\documentclass[12pt,letterpaper]{article}

\usepackage[includeheadfoot,
            marginratio={1:1,2:3}, 
            width=412pt, 
            height=688pt,]{geometry}

\usepackage{amsmath,mathrsfs}
\usepackage{amsfonts}
\usepackage{amssymb}
\usepackage{stmaryrd}
\usepackage{graphicx}
\DeclareMathAlphabet{\mathpzc}{OT1}{pzc}{m}{it}

\usepackage{empheq}
\usepackage{paralist}

\newcommand{\nc}{\newcommand}
\nc{\lb}{\llbracket}
\nc{\rb}{\rrbracket}
\nc{\gl}{\llbracket}
\nc{\gr}{\rrbracket}

\newcommand{\eq}[1]{\begin{equation}
                     \begin{split} #1 \end{split}
                     \end{equation}}

\newcommand{\ov}{\overline}

\newcommand{\eqb}[1]{\begin{equation}
                  \setlength\fboxsep{0.3cm}
                   \setlength\fboxrule{0.5pt}
                    \boxed{
                     \begin{split} #1 \end{split}}
                     \end{equation}}

\newcommand{\pd}{\partial}

\allowdisplaybreaks[2]
\numberwithin{equation}{section}

\begin{document}

\vspace*{-1.5cm}
\begin{flushright}
  {\small
  MPP-2018-15\\
 LMU-ASC 05/18\\
  }
\end{flushright}

\vspace{1.0cm}
\begin{center}
{\LARGE
Bootstrapping Non-commutative   \\[0.2cm]
Gauge Theories from L$_\infty$ algebras \\[0.2cm]
} 
\vspace{0.3cm}

\end{center}

\vspace{0.30cm}
\begin{center}
  Ralph Blumenhagen$^{1}$, Ilka Brunner$^{2}$, Vladislav Kupriyanov$^{1,3,4}$, Dieter L\"ust$^{1,2}$
\end{center}

\vspace{0.1cm}
\begin{center} 
\emph{$^{1}$ Max-Planck-Institut f\"ur Physik (Werner-Heisenberg-Institut), \\ 
   F\"ohringer Ring 6,  80805 M\"unchen, Germany } \\[0.1cm] 
\vspace{0.25cm} 
\emph{$^{2}$ Arnold Sommerfeld Center for Theoretical Physics,\\ 
               LMU, Theresienstr.~37, 80333 M\"unchen, Germany}\\[0.1cm] 
\vspace{0.25cm} 
\emph{$^{3}$ CMCC-Universidade Federal do ABC, Santo Andr\'e, SP, 
Brazil}\\[0.1cm] 
\vspace{0.25cm} 
\emph{$^{4}$ Tomsk State University, Tomsk, Russia}\\

\vspace{0.0cm}

 \vspace{0.3cm} 
\end{center} 

\vspace{0.5cm}


\begin{abstract}
Non-commutative gauge theories with a  non-constant NC-parameter
are investigated. As a novel approach, we propose that 
such theories should admit  an underlying L$_\infty$ algebra,
that governs not only the action of the symmetries but also the
dynamics of the theory.
Our approach is well motivated from string theory. We recall that
such field theories arise in the context
of branes in WZW models and briefly comment on its appearance for integrable deformations of AdS$_5$
sigma models. For the $SU(2)$ WZW model, we show 
that the earlier proposed matrix valued gauge theory on the fuzzy 2-sphere 
can be bootstrapped via an  L$_\infty$ algebra.
We then apply this approach to the construction of  non-commutative  Chern-Simons and
Yang-Mills theories on flat and curved backgrounds with non-constant
NC-structure. More concretely, up to the second order, we demonstrate how
derivative and curvature corrections to the equations of motion can be 
bootstrapped in an algebraic way  from the L$_\infty$ algebra.
The appearance of a non-trivial A$_\infty$ algebra is discussed, as well.
\end{abstract}

\clearpage

\tableofcontents



\section{Introduction}
\label{sec:intro}

It is one of the rather appealing features of string theory that
the effective theory on D-branes in a two-form background
is given by a non-commutative (NC) Yang-Mills theory \cite{Seiberg:1999vs}. For constant
two-form flux this result can be explicitly derived by quantizing the
open string and computing conformal field theory correlation
functions \cite{Schomerus:1999ug}. In this case, the non-commutative theory is governed by the
associative Moyal-Weyl star-product.

From string theory it is known that there also exist consistent D-brane
solutions of the string equations of motion that wrap curved submanifolds
and carry a non-constant two-form flux, thus  leading to a non-constant
non-commutativity structure\footnote{In this paper we use the term
non-commutativity structure for a non-constant $\Theta^{ij}$.}
$\Theta^{ij}$. Examples are branes in
WZW models \cite{Alekseev:1998mc} or holographic duals of integrable deformations
of AdS$_5$ sigma models \cite{Bena:2003wd}. In the latter case, the holographic
dual gauge theory still lives on flat space and only receives a deformation 
in the non-commutativity structure.
Therefore, one expect that
one can formulate a non-commutative gauge theory also for such more
general cases. Using techniques from conformal field theory, for the $SU(2)$ WZW model
it was shown that this theory is a non-commutative matrix valued gauge
theory on the fuzzy 2-sphere. This theory is still associative, but in
principle also this could be broken. Throughout this paper
we will be agnostic about this point and admit also 
non-associative star-products.

There have been some attempts to provide a description
of such gauge theories using the general Kontsevich \cite{Kontsevich:1997vb} star-product
\cite{Madore:2000en, Jurco:2000ja, Jurco:2001rq, Behr:2003qc} 
(for a recent application  see \cite{Chatzistavrakidis:2018vfi})
and
invoking techniques from Hopf-algebras \cite{Dimitrijevic:2003pn,Dimitrijevic:2011jg}.
For more information and literature on these attempts please  consult the review
\cite{Szabo:2006wx}.
However, these approaches were motivated rather mathematically while missing
a clear physical guiding principle for their construction.
It is the aim of this paper, to take such a physical principle from string
theory and to analyze whether it works and gives reasonable results.

We note that more recently there have also been proposals for the
appearance of non-commutative and non-associative structures in the 
closed string sector \cite{Blumenhagen:2010hj,Lust:2010iy}, in particular when one has a non-geometric flux
background. Let us emphasize that in this paper we restrict to the
open string case with D-branes.

In this paper we propose that the missing physically motivated 
guiding principle is the existence of an L$_\infty$ (or A$_\infty$)
algebra. Before  we investigate this idea in more detail,   let us mention that 
these structures appeared for the first time in the context of 
string field theory \cite{Zwiebach:1992ie}.  Indeed, e.g. for bosonic closed string field
theory, both the action of symmetries on the string field and their
string equations of motion were governed by an L$_\infty$ algebra.
The latter can be considered as a generalization of a Lie algebra,
where one allows field dependent gauge parameters. This weakens
the closure constraint and motivates the introduction of in general
infinitely many  higher 
products satisfying generalized Jacobi
identities.
These  are quadratic expressions involving for each $n$ finitely
many higher products. In particular, the usual Jacobi identity
for the two-product (the commutator) can be violated by ''derivative''
terms, thus allowing a mild form of non-associativity. For this reason,
in the mathematics literature such algebras have been  called strong
homotopy algebras \cite{Lada:1992wc}.

In \cite{Hohm:2017pnh}, the authors showed that L$_\infty$ algebras do not only
show up in string field theory, but also in much simpler field theories,
like Chern-Simons (CS) and Yang-Mills (YM) theories.
Here the structure is  considerably  truncated and  only a finite number
of higher products and relations were non-trivial. It is very
tantalizing that again not only the action of the symmetry but also
the dynamics of the whole gauge theory fit into such finite
 L$_\infty$ algebras. The authors also proposed that every consistent
gauge theory should be governed by such an underlying
L$_\infty$ algebra. 

In \cite{Blumenhagen:2017ogh,Blumenhagen:2017ulg}, motivated by the AdS$_3$-CFT$_2$ 
holographic duality, it was shown that ${\cal W}$-algebras, describing 
infinitely many global symmetries in two-dimensional conformal field 
theories, also feature an underlying, highly non-trivial  L$_\infty$ structure.
Here, it was the non-linearity of the ${\cal W}$-algebra that induced  higher 
products and relations. Turning the logic around, if they were not
already known, ${\cal W}$-algebras could have  been bootstrapped from the 
L$_\infty$ algebra.
Furthermore it was shown in \cite{Hohm:2017cey} that the non-associative closed string R-flux algebra
as well as the associated M-theory R-flux algebra of the seven
octonions can be extended to a 2-term  L$_\infty$ algebra.

Thus, so far there exist a couple of physical examples that could
be rewritten in terms of  L$_\infty$ algebras.  
The motivation for this work is to advance the symmetry concept  
of L$_\infty$ algebras and  actually exploit  it to determine the
structure of the above mentioned  NC gauge theories with general NC-structure.
For this purpose we will follow a bootstrap approach, where we
take some initial lower order products, like one- and two-products,
and bootstrap the remaining higher products by invoking the
 L$_\infty$ relations. The initial data are essentially the first term
in the gauge variation and in the equations of motion, i.e. the
one resulting from a kinetic term in the action. All of the rest follows.
For a general NC-structure, we will see that all the other
higher products receive derivative $(\partial\Theta)$-corrections. 
In other words, imposing the guiding principle 
of an underlying L$_\infty$ algebra, we can algebraically bootstrap
the derivative corrections to the action
of the  NC gauge  symmetry onto the gauge fields and their equations
of motion. 

In this paper, we explicitly show how this procedure can be carried
out in detail up to second order in $\Theta$. For this purpose, in
section 2 we review 
some facts  about NC gauge theories and recall the mathematical
definitions
of L$_\infty$ and A$_\infty$ algebras. As a first application of
L$_\infty$ algebras, we show that 
NC-CS and NC-YM theories on the Moyal-Weyl plane fit into this
scheme. 

In section 3 we remind the reader of concrete string theory settings
where NC gauge theories with non-constant NC-structure appeared.
These are branes in WZW models and holographic duals of 
integrable deformations of AdS$_5$ sigma models.
For the $SU(2)$ WZW model, 
we recall that  an NC matrix valued gauge theory has been derived via
CFT techniques \cite{Alekseev:2000fd}.
As a compelling first result, 
we show that this unconventional  NC gauge theory theory on the
fuzzy 2-sphere can be bootstrapped by imposing the existence of an L$_\infty$ algebra. 

In section 4, we apply the same technique to the more general class 
of NC gauge theory on flat and curved  space with non-constant NC-structure.
This is done in a perturbative approach in $\Theta$.
First, we point out  the essential  problem arising  for
non-constant $\Theta$ and argue that it receives a natural solution in the
context of L$_\infty$ algebras. 
We bootstrap the derivative corrections to the action of the NC gauge symmetry onto
the gauge fields. Then, we extend the L$_\infty$ algebra to also include the equations of
motion of a NC-CS and a NC-YM theory on flat and curved space.

We show that up to second order in $\Theta$ one
can even find an A$_\infty$ algebra. As expected, the graded symmetrization of the obtained structure results in the corresponding L$_\infty$ algebra. Since this involves a lengthy and
tedious  computation, we have delegated this part to an appendix - not
because it is less important but for not too much disturbing the main flow
of the paper.

\section{Preliminaries}

For self-consistency, in this section we introduce  some of the
salient features of known  NC  gauge theories and the formal
definitions of  L$_\infty$ and A$_\infty$ algebras. 
In addition, we analyze the NC gauge theory based on the Moyal-Weyl
star product with respect to an underlying L$_\infty$ algebra.

\subsection{Non-commutative gauge theories}
\label{sec_NCgauge}

First, let us recall that the conformal field theory of an open
string ending on a D-brane supporting a non-trivial 
gauge flux ${\cal F}=B+2\pi F$ features 
a non-commutative geometry.  In this paper we choose $\alpha'=1$.
Indeed, by computing the disc level scattering amplitude of
$N$-tachyons, certain relative phases appear which 
for constant gauge flux can be described by the Moyal-Weyl
star-product
\eq{
\label{starproduct}
   (f\,\star\, g)(x)= \exp\biggl(
    {i\over 2}\, \Theta^{ij}\,
      \partial^{x_1}_{i}\,\partial^{x_2}_{j} \biggr)\, f(x_1)\, g(x_2)\Bigr|_{x} \;.
}
The open string quantities governing the theory on the D-brane are 
related to the initial closed string variables $g$ and ${\cal F}$ via
 $G^{-1}+\Theta=(g+{\cal F})^{-1}$, where
the anti-symmetric bi-vector $\Theta^{ij}$ is the one appearing 
in the star product \eqref{starproduct}.
In the Seiberg-Witten limit the OPE exactly becomes the 
Moyal-Weyl star-product.
This non-trivial product of functions
leads to the non-commutative Moyal-Weyl plane with
$[x^i,x^j]_\star=i\, \Theta^{ij}$.
In \cite{Kontsevich:1997vb}  it  has been shown that for every Poisson
structure $\Theta^{ij}$ that by definition satisfies
\eq{
                        \Pi^{ijk}:=3\,\Theta^{[ im} \partial_m
                        \Theta^{jk]}=0\,
}
one can define a corresponding  associative star-product, which 
will also involve derivatives of the Poisson structure.
The same product can also be considered for a quasi Poisson structure,
but then leads to a non-associative star-product, which up to second
order in $\Theta$ reads
\eq{
\label{starkont}
     f\bullet g=f\cdot g + 
  &{i\over 2} \Theta^{ij}\, \partial_i f\, \partial_j g -
   {1\over 8} \Theta^{ij}\Theta^{kl}\, \partial_i\partial_k f\, \partial_j\partial_l g  \\
  &-{1\over 12} \big(\Theta^{im}\partial_m \Theta^{jk}\big) \big(
  \partial_i\partial_j f \, \partial_k g + \partial_i\partial_j g\, \partial_k f\big) + O(\Theta^3)\, .
}
For the higher order expression see  \cite{Penkava, Herbst:2003we, Kupriyanov:2008dn, Kupriyanov:2015dda}.
Often we will write this as
\eq{
     f\bullet g=f\star g
  -{1\over 12} \big(\Theta^{im}\partial_m \Theta^{jk}\big) \big(
  \partial_i\partial_j f \, \partial_k g + \partial_i\partial_j g\, \partial_k f\big) + O(\Theta^3)\, 
}
that separates the derivative $\partial \Theta$-corrections from the
standard Moyal-Weyl terms. 
The associator for this product becomes 
\eq{
\label{openasso}
    (f\bullet g)\bullet h-f\bullet (g\bullet h)={1\over 6}\,
      \Pi^{ijk}\, \partial_i f \,\partial_j g \,\partial_k h+
      O(\Theta^3)\, .
}
In \cite{Seiberg:1999vs}, for  the Moyal-Weyl case with constant open string metric
and NC-parame\-ter,  it was shown that the effective
theory on a stack of $N$ branes is given by a non-commutative gauge
theory with gauge group $U(N)$. In the following we stick to the
$U(1)$ case.  As in usual YM theory there is a gauge field $A_a(x)$
behaving under a gauge transformation as
\eq{
\label{gaugevara}
        \delta_f A_a =\partial_a f +i [f,A_a]_*\,.
}
Using the Leibniz rule for the star-bracket $[.,.]_\star$ and
its associativity\footnote{Note that both assumptions might not be
  satisfied  for non-constant $\Theta$.}, one can show  that two gauge
transformations close off-shell in the sense 
\eq{
\label{gaugeclosea}
            [\delta_f,\delta_g] A_a= \delta_{-i[f,g]_\star} A_a\,.
}
Moreover, the field-strength
\eq{
              F_{ab}=\partial_a A_b-\partial_b A_a-i[A_a,A_b]_\star
}
transforms covariantly, i.e.
\eq{
                    \delta_f F_{ab}= i [f,F_{ab}]_\star\,.
}
Then, the vacuum equation of motion for the non-commutative $U(1)$ Yang-Mills theory reads
\eq{
\label{YM3}
0&={\mathpzc F}_a=\partial^b F_{b a}-i[A^b,F_{ba}]_\star\\[0.2cm]
&=\Box A_{a}-\partial_a(\pd\cdot A)-i\,\partial^b[A_b,A_a]_\star-i[A^b,\partial_b
A_a-\partial_a A_b]_\star\\[0.1cm]
&\phantom{=}-[A^b,[A_b,A_a]_\star]_\star\; .
}
In section \ref{sec_CSYML} we will come back to these relations and study their
implementation into an L$_\infty$ algebra.

Similarly one can also define a non-commutative Chern-Simons theory in
three-dimensions, whose equation of motion is
\eq{
\label{NCCSeom}
       0={\mathpzc F}_c=\epsilon_c{}^{ab} \Big(\partial_a A_b -{i\over 2}
       [A_a,A_b]_\star \Big)\,.
}
For usual CS and YM-theory, it was explicitly shown in \cite{Hohm:2017pnh} that both their
symmetries and their  dynamics are governed in an algebraic way by the
objects and relations of an L$_\infty$ algebra. Before  reviewing this,
in the next section we give a brief introduction into the general
notion of  L$_\infty$ and also of A$_\infty$ algebras.

\subsection{L$_\infty$ and A$_\infty$ algebras and gauge symmetries}
\label{sect_22}

Following \cite{Hohm:2017pnh}, let us review the basis notion of L$_\infty$ and A$_\infty$ algebras
and the generic relation of the first to the description of gauge
symmetries and their dynamics.

\subsubsection*{Definition of  L$_\infty$ algebra}

L$_\infty$ algebras are generalized Lie algebras where one has not only a two-product, the commutator, but more general multilinear $n$-products with $n$ inputs
\eq{
\ell_n: \qquad \quad X^{\otimes n} &\rightarrow X \\
x_1, \dots , x_n &\mapsto \ell_n(x_1, \dots , x_n) \, , 
}
defined on a graded vector space $X = \bigoplus_n X_n$, where $n$ denotes the grading.
These products are graded anti-symmetric 
\eq{ 
\label{permuting}
\ell_n (\dots, x_1,x_2, \dots) = (-1)^{1+ {\rm deg}(x_1) {\rm deg}( x_2)} \, \ell_n (\dots, x_2,x_1, \dots )\,,
}
with 
\eq{
      {\rm deg}\big( \, \ell_n(x_1,\ldots,x_n)\, \big)=n-2+\sum_{i=1}^n  {\rm deg}(x_i)\,.
}
 The set of higher products  $\ell_n$ define an L$_\infty$ algebra, if they satisfy the
 infinitely many relations
\eq{
\label{linftyrels}
{\cal J}_n(x_1,\ldots, x_n):=\sum_{i + j = n + 1 } &(-1)^{i(j-1)}
\sum_\sigma  (-1)^\sigma
\, \chi (\sigma;x) \; \\
 &\ell_j \big( \;
\ell_i (x_{\sigma(1)}\; , \dots , x_{\sigma(i)} )\, , x_{\sigma(i+1)} , \dots ,
x_{\sigma(n)} \, \big) = 0 \, .
}
The permutations are restricted to the ones with
\eq{ 
\label{restrictiononpermutation}
\sigma(1) < \cdots < \sigma(i) , \qquad \sigma(i+1) < \cdots < \sigma(n)\,,
}
and the sign $\chi(\sigma; x) = \pm 1$ can be read off from \eqref{permuting}. 
The first relations ${\cal J}_n$ with $n=1,2,3,\ldots$ can be schematically written as 
\eq{ 
&{\cal J}_1 = \ell_1 \ell_1 \, , \qquad {\cal J}_2 = \ell_1 \ell_2 - \ell_2 \ell_1 \,,  \qquad {\cal J}_3 =
\ell_1 \ell_3 + \ell_2 \ell_2 + \ell_3 \ell_1 \,, \\[0.1cm] &
 {\cal J}_4 = \ell_1 \ell_4 - \ell_2 \ell_3 + \ell_3 \ell_2 - \ell_4 \ell_1 \, , 
}
from which one can deduce the scheme for the higher ${\cal J}_n$. More concretely, the first L$_\infty$ relations read
\eq{   \label{ininftyrel1}
\ell_1\big( \,  \ell_1   (x) \, \big) &= 0 \\
\ell_1 \big( \, \ell_2(x_1, x_2)\,\big) &=  \ell_2\big(\,  \ell_1 (x_1) , x_2 \, \big) + (-1)^{x_1} \ell_2\big(\, x_1, \ell_1 (x_2) \, \big) \, ,
}
revealing that $\ell_1$ must be a nilpotent derivation with respect to
$\ell_2$, i.e. that in particular the Leibniz rule is satisfied.
Denoting $(-1)^{x_i}=(-1)^{deg(x_i)}$ the full relation $ {\cal J}_3 $ reads
\begin{eqnarray}    \label{ininftyrel2}
       0\!\!\!&=&\!\!\! \phantom{} \ell_1\big(\ell_3(x_1,x_2,x_3)\, \big) +\\[0.2cm]
     &&\ell_2\big(\ell_2(x_1,x_2),x_3\, \big)+(-1)^{(x_2+x_3)x_1}
     \ell_2\big(\ell_2(x_2,x_3),x_1\, \big)+\nonumber \\[0.1cm]
   &&(-1)^{(x_1+x_2)x_3}
     \ell_2\big(\ell_2(x_3,x_1),x_2\, \big)+\nonumber \\[0.2cm]
     &&\ell_3\big(\ell_1(x_1),x_2,x_3\, \big)+(-1)^{x_1}\ell_3\big(x_1,\ell_1(x_2),x_3\, \big)
+(-1)^{x_1+x_2}\ell_3\big(x_1,x_2,\ell_1(x_3)\, \big)\, \nonumber
\end{eqnarray}   
and means that 
the Jacobi identity for the $\ell_2$ product is mildly violated by
$\ell_1$ exact expressions.

\subsubsection*{Definition of A$_\infty$ algebras}

The definition of an A$_\infty$ algebras is very similar to the
definition of an L$_\infty$ algebra. While L$_\infty$ algebras are generalized differential graded Lie algebras with a mild violation of the Jacobi identity, A$_\infty$ algebras generalize algebras with a mild violation of associativity. One has higher products
$m_n(x_1,\ldots, x_n)$ of degree $n-2$, where $x_i$ are again elements of a graded vector space. The quadratic relations for the higher
products are
\eq{
     {\cal A}_{n-1}(x_1,\ldots,x_{n-1})=\sum_{l=1}^{n-1}    (-1)^ {n(l+1)} \, m_l \circ m_{n-l}=0
}
with the second product  defined as
\eq{
                 m_p=\sum_{r=0}^{n-1-p}  (-1)^{r(p+1)}\, 1^r \otimes
                 m_p\otimes 1^{n-1-p-r}\,.
}
The first three relations read
\eq{
\label{relaerst}
          {\cal A}_1 &= m_1 \circ  m_1\\
           {\cal A}_2 &= m_1 \circ  m_2 - m_2\circ  (m_1\otimes 1 +
           1\otimes m_1)\\
 {\cal A}_3 &= m_1 \circ  m_3 + m_2\circ  (m_2\otimes 1 - 1\otimes m_2)\\
&\phantom{=} +m_3 \circ  ( m_1\otimes 1 \otimes 1
+ 1\otimes m_1 \otimes 1+ 1\otimes 1 \otimes m_1)
}
where whenever an  odd degree $m_n$ is exchanged with an odd degree $x_m$  one gets an extra
minus sign. This will become clearer in appendix \ref{app_B} where
we will consider explicit examples.
There we also need the next relation
\eq{
\label{Arelationfour}
{\cal A}_4=\,&m_1\circ  m_4-
m_2\circ ( m_3\otimes 1+ 1\otimes m_3)\\
&+m_3\circ  ( m_2\otimes 1 \otimes 1
- 1\otimes m_2 \otimes 1+ 1\otimes 1 \otimes m_2)\\
&-m_4\circ (m_1\otimes 1^3 +1\otimes m_1 \otimes
1^2+1^2 \otimes m_1\otimes 1+1^3\otimes m_1)\,.
}
Even though gauge theories arise for the open string and string field
theory suggest that they are related by an A$_\infty$ structure, 
\cite{Hohm:2017pnh} proposed that they also fit nicely into the structure of 
L$_\infty$ algebras.

\subsubsection*{Gauge theories and L$_\infty$ algebras}
\label{sec_gaul}

The framework of L$_\infty$ algebras is quite flexible
and it has been suggested  that every classical perturbative gauge
theory (derived from string theory),
including its dynamics, is organized by an underlying L$_\infty$
structure \cite{Hohm:2017pnh}. For sure, the pure gauge algebra,
called L$_\infty^{\rm gauge}$,  of such theories satisfies
the L$_\infty$ identities.  To see this, let us assume that the field theory has a standard
type gauge structure, meaning that the variations of the fields can be
organized unambiguously into a sum of terms each of a definite power
in the fields. 
Then we choose only two non-trivial vector spaces as
\eq{      
\begin{matrix}   X_0\quad  &\quad X_{-1} \\[0.2cm]
                        f    \quad  &\quad   A_a  
\end{matrix}\;.
\label{fA}
}
In this case, the only allowed non-trivial higher product are the ones
with one and two gauge parameters $\ell_{n+1}(f, A^n )\in X_{-1}$ and
$\ell_{n+2}(f, g, A^n )\in X_0$ and the only non-trivial relations 
are ${\cal J}_{n+2}(f, g, A^n )\in X_{-1}$ and  ${\cal J}_{n+3}(f, g,h,
A^n )\in X_{0}$.
Then, the gauge variations are expanded as
\eq{\label{var}
  \delta_{f}  A &=\sum_{n\ge 0}   {1\over n!}
      (-1)^{n(n-1)\over 2}\,
 \ell_{n+1}(f, \underbrace{ A, \dots, A}_{n \; {\rm times}} )\, .
 }
This allows to read off the higher products $\ell_{n+1}(f,A^{n})\in X_{-1}$.
It was shown in
\cite{Berends:1984rq,Fulp:2002kk,Hohm:2017pnh}, that the
off-shell closure of the symmetry variations 
\eq{
\label{commurel4}
                      [\delta_{f},\delta_g] A
                      =\delta_{- C(f,g, A)} A \, ,
}
and the Jacobi identity
 \eq{ \label{Gaugejacobiator4}
\sum_{\rm cycl} \big[ \delta_{f}, [  \delta_{g} ,  \delta_{h} ] \big] = 0 \,  
}
are equivalent to the L$_\infty$ relations with two and three gauge parameters.
Here the  closure relation allows for a field dependent gauge parameter which can be written in terms of L$_\infty$ products as
\eq{\label{clclosure}
C(f,g, A) & =\sum_{n\ge 0} {1\over n!}
            (-1)^{{n(n-1)\over 2}}
            \, \ell_{n+2}(f,g,  \underbrace{ A, \dots, A}_{n\;  {\rm times}})\,.
} 
Thus, the action of gauge symmetries on the fundamental fields is
governed by an L$_\infty^{\rm gauge}$ algebra. However, this is not
the end of the story, as string field theory suggests that also
the dynamics of the theory, i.e. the equations of motion, are expected
to fit into an extended L$_\infty^{\rm full}$ algebra.

For this purpose one extends the vector space to $X_0\oplus X_{-1}
\oplus X_{-2}$ 
\eq{      
\begin{matrix}   X_0\quad  &\quad X_{-1}\quad  &\quad  X_{-2} \\[0.2cm]
                        f    \quad  &\quad   A_a   \quad  &\quad  E_a
\end{matrix}\label{fAE}
}
where $X_{-2}$ also contains the equations of motion, 
i.e. ${\mathpzc F}\in X_{-2}$.
Now many more higher products can be non-trivial and one has to check in a
case by case study whether indeed the  L$_\infty^{\rm full}$ algebra
closes. 
The higher products $\ell_{n}(A^{n})\in X_{-2}$ are special as
they give the equation of motion that is expanded as
\eq{
{\mathpzc F}=\underset{n \ge 1}{\sum} \;{1\over n!}
(-1)^\frac{n(n-1)}{2}\, \ell_n(A^n)=
   \ell_1(A)-{1\over 2} \ell_2(A^2)-{1\over 3!} \ell_3(A^3)+\ldots\; .
}
Moreover, the structure admits that the closure condition \eqref{commurel4}
is only satisfied on-shell, i.e. there can be terms $\ell_{n+3}
(f,g,{\mathpzc F},A^n)\in X_{-1}$
on the right hand side. In case one has off-shell closure (like for
the CS and YM theories considered in this paper) all these higher
product are vanishing.
Moreover, the gauge variation of ${\mathpzc F}$ is given by
\eq{
\label{varF}
  \delta_{f}  {\mathpzc F} &=\ell_2(f,{\mathpzc F})+ \ell_3(f,{\mathpzc F},A)-{1\over 2} \ell_4(f,{\mathpzc F},A^2)+\ldots
}
reflecting that, as opposed to the gauge field $A$, it transforms covariantly.       

It was proposed that for  writing down an action for these
equations
of motion one needs an  inner product
\eq{
                 \langle \ \, , \ \rangle: X_{-1}\otimes X_{-2}\to
                 \mathbb R
}
satisfying the  cyclicity property
\eq{
\label{cyclicprop}
                \langle A_0,\ell_n(A_1,\ldots,A_n) \rangle=  \langle A_1,\ell_n(A_0,\ldots,A_n )\rangle\,
}     
for all $A_i\in X_{-1}$.          
Then, the equations of motion follow from varying the action
\eq{
\label{actionA}
                   S&=\underset{n \ge 1}{\sum} \;{1\over (n+1)!}
(-1)^\frac{n(n-1)}{2}\, \langle A,\ell_n(A^n)\rangle\\[0.1cm]
&=  {1\over 2}\langle A,\ell_1(A)\rangle-{1\over 3!}  \langle A, \ell_2(A^2)\rangle-{1\over 4!}
    \langle A, \ell_3(A^3)\rangle+\ldots\; .
}

\subsection{L$_\infty$ algebras for NC-CS and NC-YM gauge theories} 
\label{sec_CSYML}

Now, as two examples, we analyze how $U(1)$ NC Chern-Simons and Yang-Mills
theories fit into the scheme of L$_\infty$ algebras. In this section,
we consider the case of the Moyal-Weyl star-product, i.e. the
NC-parameter $\Theta$ is constant. In this case, the computation is 
very similar to the analysis of ordinary (non-abelian) CS and YM
theories discussed in \cite{Hohm:2017pnh}. 

\subsubsection*{L$_\infty$ structure of non-commutative CS} 

The vector spaces are still as in eq.(\ref{fA}) or as in eq.(\ref{fAE}).
Some of the relevant relations have already been given in section
\ref{sec_NCgauge}.
From the gauge variation \eqref{gaugevara}, we can read off
\footnote{Note that by writing $\ell_2(f,A)$ it is understood that 
the object also carries an index $a$ like $\ell_2(f,A)_a$. In order
not to clutter the notation, in the following we leave this index out,
as it is usually clear from the free index on the r.h.s..}
\eq{
\label{NCCS1}
     \ell_1(f)=\partial_a f\,,\qquad \ell_2(f,A)=i[f,A_a]_\star
}
and from the off-shell closure condition \eqref{gaugeclosea}
\eq{
\label{NCCS2}
          \ell_2(f,g)=i [f,g]_\star\,, \qquad \ell_3(f,g,E)=0\,
}
with all higher products vanishing, e.g. $ \ell_{n+1}(f,A^n)=0$ for
$n\ge 2$.
The equation of motion \eqref{NCCSeom} motivates the choice for   the non-vanishing products
\eq{
                  \ell_1(A)=\epsilon_c{}^{ab} \partial_a A_b
                  \,,\qquad
                  \ell_2(A,B)=i\epsilon_c{}^{ab} [A_a, B_b]_\star\,.
}
Therefore, only $\ell_1$ and $\ell_2$ products are non-vanishing and
one only has to check the  finite number of L$_\infty$ relations
listed below
\eq{
\label{relasCS}
           &{\cal J}_1(f) \in X_{-2} \\[0.1cm]
           & {\cal J}_2(f,g) \in X_{-1}\,,  \quad\ \, {\cal J}_2(f,A)\in X_{-2} ,\\[0.1cm]
          & {\cal J}_3(f,g,h) \in X_{0} \,,  \quad {\cal J}_3(f,g,A)\in X_{-1} 
\,,  \quad {\cal J}_3(f,A,B)\in X_{-2} \\
  &  {\cal J}_3(f,g,E)\in X_{-2} \,.  
}
The first relation ${\cal
  J}_1(f)=\ell_1(\ell_1(f))=\epsilon_c{}^{ab} \partial_a \partial_b f=0$  can be readily checked.
The relation ${\cal J}_2(f,g)=0$ is nothing else than the Leibniz-rule
for the star commutator. The full third relation reads
\eq{
          \ell_1(\ell_2(f,A))=\ell_2(\ell_1(f),A)+ \ell_2(f,\ell_1(A))
}
which fixes the last term to be 
\eq{
\ell_2(f,E)=i[f, E_a]_\star\,.
 }
Since all $\ell_3$ are vanishing, the remaining four ${\cal J}_3$
relations do only contain $\ell_2\ell_2$-terms. As the star commutator
satisfies the Jacobi identity,
these are all satisfied.
Let us also mention that the field strength can be expressed as
\eq{
                 \ell_1(A)-{1\over 2} \ell_2(A,A)={1\over 2}\epsilon_c{}^{ab}
                 \Big(\partial_a A_b -\partial_b A_a -i[A_a,A_b]_\star\Big)={1\over 2}
                 \epsilon_c{}^{ab} F_{ab}\,.
}
Clearly, by setting all elements in $X_{-2}$ to zero, one gets the sub-algebra
L$_\infty^{\rm gauge}$. The latter is the same for NC-CS and NC-YM.
Defining the inner product as
\eq{
            \langle A,E\rangle=\int d^3 x   \;\eta^{ab}  A_a E_b 
}
one can integrate this to an action \eqref{actionA}.

Thus, we have seen that the $U(1)$ NC-CS theory fits nicely into the 
L$_\infty$ framework, where the highest appearing products are
$\ell_2$.

\subsubsection*{L$_\infty$ structure of non-commutative YM}

A similar computation can also be done for NC-YM theory. The case
of usual non-abelian YM-theory was first formulated in
\cite{Zeitlin:2007vv, Zeitlin:2008cc}. 
Here we follow the same path as in the more recent paper
\cite{Hohm:2017pnh}.

Since the action of a gauge transformation on the fields and its
closure are the same as for NC-CS theory, the products $\ell_1(f)$ and
$\ell_2(f,A)$ from \eqref{NCCS1} and $\ell_2(f,g)$ from  \eqref{NCCS2}
are still valid.
The equations of motion \eqref{YM3} allow one to read-off the higher
  products
\eq{
     \ell_1(A)&=\Box A_{a}-\partial_a(\pd\cdot A) \\[0.1cm]
    \ell_2(A,B)&=i \,\partial^b [A_b,B_a]_\star +i [A^b,\partial_b
    B_a-\partial_a B_b]_\star +(A \leftrightarrow B) \\[0.1cm]
    \ell_3(A,B,C)&=[A^b,[B_b,C_a]_\star ]_\star + [B^b,[C_b,A_a]_\star ]_\star
                                   + [C^b,[A_b,B_a]_\star ]_\star\\
                                   &+ [A^b,[C_b,B_a]_\star ]_\star + [C^b,[B_b,A_a]_\star ]_\star
                                   + [B^b,[A_b,C_a]_\star ]_\star\,.
}
Note that $\ell_1(A)$ has changed from the NC-CS case and that for NC-YM there also exist a non-vanishing
$\ell_3$. Therefore, besides \eqref{relasCS} one also has to check the
L$_\infty$ relations
\eq{
\label{identiorderfour}
   & {\cal J}_4(f,g,h,A) \in X_{0} \,,  \quad {\cal J}_4(f,g,A,B)\in X_{-1} 
\,,  \quad {\cal J}_4(f,g,h,E)\in X_{-1} \\[0.1cm]
  &  {\cal J}_4(f,A,B,C)\in X_{-2}  \,,  \quad {\cal J}_4(f,g,A,E)\in X_{-2} \,.  
}
The nil-potency condition $\ell_1(\ell_1(f))=0$ can readily be checked.
Similarly to the NC-CS theory, the Leibniz-rule ${\cal J}_2(f,A)$
fixes $\ell_2(f,E)=i[f, E_a]_\star$. Setting now all other higher
products to zero, one realizes that the three relations
 ${\cal J}_3(f,g,h)={\cal J}_3(f,g,A)={\cal J}_3(f,g,E)=0$ involve
only star-commutators and are satisfied by their Jacobi identity.
The identity ${\cal J}_3(f,A,B)=0$ is more non-trivial and also involves
the three-product $\ell_3(A,B,C)$. However, by spelling out all terms
in the relation, one realizes that  they indeed all cancel.
From the next order relations in \eqref{identiorderfour} only
$ {\cal J}_4(f,A,B,C)=0$ is non-trivial, but can be checked by
applying the Jacobi identity for the star-commutator.
In principle also $ {\cal J}_5$ could be relevant, but due to 
$\ell_3(\ell_3(A^3),A^2)\in X_{-3}$ these relations are satisfied trivially.

\section{NC gauge theories arising in string theory}
\label{sec_Dbranes}

We just showed that  both NC-CS and NC-YM on flat Minkowski space with
constant NC-structure 
$\Theta$ fit into the scheme of L$_\infty$ algebras. However, not all
consistent D-branes (boundary states) in string theory are of this
simple type, as there do also exist D-branes wrapping curved submanifolds 
and carrying a non-constant gauge flux on the  brane world-volume.
Therefore, the question arises whether also the expected NC gauge
theory on such branes fit into the scheme of L$_\infty$ algebras. In
this case, the Kontsevich star product \eqref{starkont} indicates that one
gets extra derivative terms $\partial \Theta$. 

Before we continue along these lines, in this section we want to 
remind the reader of a few  stringy circumstances  where 
non-commutativity with non-constant $\Theta$ does appear.
These will be branes in exactly solvable WZW models for  compact
groups and recent advances related to integrable  deformations
of AdS$_5$ sigma models.

\subsection{D-Branes in WZW models}

In this section we review some of the features of
D-branes\footnote{Even though we are working with
the bosonic string and there are no R-R fields, we call these branes
D-branes, as their tension $T\sim g_s^{-1}$ scales with
string coupling in the same manner as for D-branes in type II string
theory.}  in WZW models relevant for us. WZW models are exactly solvable sigma models
whose target spaces are group manifolds equipped with  non-trivial
NS-NS three-form fluxes. Their distinctive feature is that the corresponding  two-dimensional
conformal field theories are  explicitly known and given by the unitary series of
Kac-Moody algebras. As a consequence it was possible to construct
boundary states in the CFT that turned out to correspond to certain branes
wrapping conjugacy classes of the group manifold
\cite{Alekseev:1998mc,Alekseev:1999bs} carrying non-constant two-form flux.
In this section, we will review the semi-classical description of
these consistent branes.

\subsubsection*{Preliminaries}

The starting point is the two-dimensional world-sheet action of a WZW model
\eq{
            S_{\rm WZW}&={k\over 16\pi} \int_{\partial \Sigma}
            d^2\sigma \,  {\rm
              Tr} \left( \partial_i h^{-1} \partial^i h
            \right) \\[0.1cm]
              &\phantom{=} +{k\over 24\pi} \int_\Sigma d^3\tilde\sigma \,\epsilon^{\tilde i
              \tilde j \tilde k} \,{\rm Tr} \left(
              (h^{-1} \partial_{\tilde i} h)(h^{-1} \partial_{\tilde j} h)(h^{-1} \partial_{\tilde k} h)
            \right)
}
where $h$ denotes the general  element of a (simple) Lie-group ${\cal
  G}$ and $\Sigma$ a three-manifold whose boundary is the closed string world-sheet.
From the WZW sigma model action one can directly read off the metric
\eq{
\label{metrictot}
           g={k\over 2}\, {\rm Tr} \left( dh^{-1} \otimes dh \right)
}
and the NS-NS three-form flux
\eq{
\label{hfluxtot}
           H={k \over 6}\, {\rm Tr} \Big( (h^{-1} dh)\wedge (h^{-1} dh)
           \wedge (h^{-1} dh)\Big)\,.
}
Here the total derivative is with respect to the target space
coordinates. Since this gives a CFT, the metric and the $H$-flux
satisfy the string equations of motion for the metric and the
$B$-field at any power in $\alpha'$, it only needs some additional
input to also satisfy the dilaton equation of motion. This can be a
linear dilaton $\varphi(z)$ depending on an orthogonal direction (like it appears
for the deep throat  limit of the NS5-brane solution).

The question which D-branes can be consistently introduced into
these closed string backgrounds has been under intensive
investigation. Here we just focus on the most simple set of such
branes. Since the WZW model describes a background with a non-trivial
$B$-field, three issues arise. 

First, one expects that the effective
theory for the gauge field on the brane becomes non-commutative
with the non-commutativity being controlled  by
an  antisymmetric bi-vector
$\Theta=\Theta^{ij} \partial_i\wedge \partial_j$, which is part of
the so-called open string fields, 
\eq{
\label{openframe}
          &G=   g-{\cal F}\, g^{-1} \, {\cal F}\,, \qquad
          \Theta=({\cal F}-g\, {\cal F}^{-1}\, g)^{-1}\,,\\
           &e^{-2 \phi} \sqrt{G}= e^{-2\varphi} \sqrt{g}\,.
}
Here ${\cal F}=B+2\pi F$ (with $\alpha'=1$ and $F=dA$) denotes the gauge
invariant open string two-form and $G$ and $\phi$
are the open string metric and dilaton.

Second, the gauge field $A$ on the 
brane provides new degrees of freedom that are also governed
by equations of motion. Varying the Dirac-Born-Infeld action
with respect to $A$, one arrives at 
\eq{
\label{openeom}
              0= \partial_i \Big(e^{-\varphi} \sqrt{g+{\cal F}}\,
              \Theta^{ij}\Big)=\partial_i \Big(e^{-\phi}
              \sqrt{G}\, \Theta^{ij}\Big)\,,
}
where the indices $i,j$ are along the brane world-volume.
Since the DBI action is established only for adiabatic field configurations, there will
presumably be higher derivative corrections to this field equation.
However, for constant ${\cal F}$ it includes all higher order
$\alpha'$ corrections.

Third, due to the non-trivial $H$-flux in the bulk, its restriction
on the brane has to satisfy $d{\cal F}=H$, i.e. it is a total derivative of
a globally defined two-form. Therefore, $H$ must be trivial in the
cohomology on the brane $[H\vert_{\rm D}]=0$. This is also
called the Freed-Witten anomaly cancellation condition \cite{Freed:1999vc}.
Note that this does not
mean that $H \vert_{\rm D}$ has to vanish identically on the brane.

\subsubsection*{Geometry of D-branes in WZW models}

In this section we provide a set of branes for which the 
geometric semi-classical identification is known
\cite{Felder:1999ka,Bachas:2000ik,Pawelczyk:2000ah,Maldacena:2001xj}.
Our presentation
follows the appendix of  \cite{Maldacena:2001xj}.
Indeed in the CFT there exist boundary states corresponding to branes wrapping 
conjugacy classes 
\eq{
\label{braneconjugacy}
                   {\cal O}(h):= \{ k^{-1} h k, {\rm for\ all}\ k\in
                   {\cal G} \}\,.
}
Here, for  the element $h$ one can always choose a representative from
the Cartan torus $M(\chi)=\exp(i \chi \cdot  H)$. Then the position of
the brane is labelled by $\chi$ and the coordinates along the D-brane can
be parametrized by angular variables $\psi$ according to
\eq{
                g=N(\psi)^{-1}\, M(\chi)\, N(\psi)\,.
}
This provides suitable coordinates for this brane configuration that 
admits a very explicit description of the geometry and the fluxes.
Note that the (generic) dimension of these branes is $d={\rm dim}{\cal G}-{\rm
rk}{\cal G}$.
Since these configurations do correspond to boundary states in the
CFT, one expects that the open string equation of motion and the
Freed-Witten anomaly condition are satisfied. 
Now one defines one-forms $\theta^\alpha=\theta^\alpha{}_i \, d\psi^i$ on the D-brane via 
\eq{
\label{oneforms}
               dN N^{-1}=\theta^\alpha\, E_\alpha -
               \theta^{\ov\alpha}\, E_{\ov\alpha} +i \rho^i H_i
}
where the generators in the Cartan-Weyl basis are normalized as
${\rm Tr}(H_i\, H_j)=\delta_{ij}$ and ${\rm Tr}(E_\alpha\, E_{\ov\beta})=\delta_{\alpha\beta}$.
As usual, the dual vector-fields are given by $\hat\theta_\alpha=\hat\theta_\alpha{}^i
\, \partial_i $ with $\hat\theta=(\theta^{-1})^T$.
Then one can show that  the metric \eqref{metrictot} restricted to the brane
world-volume can be expressed in the nice way
\eq{
\label{metricbrane}
                 g\vert_{\rm D}=2k \sum_{\alpha>0} \sin^2\Big({\alpha\cdot
                   \chi\over 2}\Big) \Big(\theta^\alpha\otimes \theta^{\ov\alpha}+\theta^{\ov\alpha}\otimes \theta^{\alpha}\Big)
}
where the sum is over all positive roots. This form neatly shows the
separation  of
the dependence on the brane positions $\chi$ and the angular coordinates
$\psi$ along the brane.
Moreover, one can choose a gauge so that the NS-NS two-form has legs only
along the brane. Indeed, Choosing
\eq{
\label{bfield_brane}
                 B=-i k \sum_{\alpha>0} \Big(\alpha\cdot\chi -\sin({\alpha\cdot
                   \chi})\Big) \Big(\theta^\alpha\otimes \theta^{\ov\alpha}-\theta^{\ov\alpha}\otimes \theta^{\alpha}\Big)
}
gives $H=dB$. Therefore, the restriction of the B-field onto the brane
is also given by this expression, i.e. $B\vert_{\rm D}=B$. This by
itself does not satisfy the open string equation motion \eqref{openeom}, but has to be
supplemented by a non-trivial gauge flux on the D-brane. This is also
known quite explicitly as
\eq{
\label{gaugeflux}
                  F={i k\over 2\pi} \sum_{\alpha>0} \Big(\alpha\cdot\chi \Big) \Big(\theta^\alpha\otimes \theta^{\ov\alpha}-\theta^{\ov\alpha}\otimes \theta^{\alpha}\Big)\,.
}
The quantization of the gauge flux fixes $\chi=2\pi (\lambda
+\rho)/k$, where $\rho=\sum_{\alpha>0} \alpha/2$ denotes the
Weyl-vector and $\rho$ an element from the weight-lattice. 
Thus, the total two-form flux on the brane is given by
\eq{
                 {\cal F}=B\vert_{\rm D}+2\pi F=i k \sum_{\alpha>0} \sin({\alpha\cdot
                   \chi}) \Big(\theta^\alpha\otimes \theta^{\ov\alpha}-\theta^{\ov\alpha}\otimes \theta^{\alpha}\Big)\,.
}
Now, one can explicitly compute similar expressions for the fields
in the open string frame \eqref{openframe}. For the metric we find the
simple result
\eq{
                G=2k \sum_{\alpha>0}  \Big(\theta^\alpha\otimes \theta^{\ov\alpha}+\theta^{\ov\alpha}\otimes \theta^{\alpha}\Big)
}
and for the anti-symmetric bi-vector
\eq{
\label{bivector}
                 \Theta={i k\over 2} \sum_{\alpha>0} \cot \Big({\alpha\cdot
                   \chi\over 2}\Big) \Big(\hat\theta_\alpha\otimes \hat\theta_{\ov\alpha}-\hat\theta_{\ov\alpha}\otimes \hat\theta_{\alpha}\Big)\,.
}
Note that in the last expression the dual one-vectors
$\hat\theta_\alpha$ appear.
For the dilaton in the open string frame one gets 
\eq{
              e^{-2\phi}=e^{-2 \varphi(z)} \prod_{\alpha>0} \sin^2\Big({\alpha\cdot
                   \chi\over 2}\Big) \,,
}
which does not depend on the coordinates along the brane.
As a consequence, the open string equation of motion
\eqref{openeom} is equivalent to 
\eq{
\label{eomcovar}
              \nabla_i \Theta^{ij}=0
}
which involves the Levi-Civita connection with respect to the open string
metric $G$.

\subsubsection*{Example: $SU(2)$ WZW}

Let us discuss the most familiar case of the $SU(2)$ WZW model. 
In this case the target space is an $S^3$ with a
non-constant $H$-flux through it. Due to the Freed-Witten anomaly condition, it is clear that
there does not exist a $D$-brane wrapping the entire $S^3$.
However, a class of $D$-branes is given by
the orbit ${\cal O}(D):= \{ k^{-1} D k\}$ with $D$ denoting an element
from the one-dimensional Cartan torus. Therefore,  generically
this describes a brane wrapping a two-dimensional submanifold
of $S^3$. 

To apply the construction from the last section, we introduce a basis 
of correctly normalized generators of $SU(2)$:
$H={1\over \sqrt{2}} \sigma_3$ and
$ E_{\alpha(\ov\alpha)}={1\over {2}} (\sigma_1 \pm i \sigma_2)$ 
with the positive root $\alpha=\sqrt{2}$. Here $\sigma_i$ denote  the Pauli matrices.
The Cartan torus $D(\chi)=\exp(i \chi H)$ is 
\eq{
D(\chi)=\left(\begin{matrix}  e^{i {\chi\over \sqrt 2}}  & 0 \\
                   0 & e^{-i {\chi\over \sqrt 2}} \end{matrix}\right)
}
and the orthogonal directions to the D-brane can be parametrized by
\eq{
N(\varphi,\psi)=\left(\begin{matrix}  \cos \varphi & \sin \varphi
    \, e^{i \psi}  \\
          -\sin \varphi \, e^{-i \psi}  & \cos \varphi 
          \end{matrix}\right)\,
}
so that we write an element of $SU(2)$ as $M= N^{-1}  D( \chi) N$.
Evaluating \eqref{metrictot}, the  metric on the $SU(2)$ group manifold reads
\eq{
\label{SU2metric}
           k^{-1}\,  ds^2={1\over 2} d\chi^2 + 4\sin^2\big({\textstyle{\chi\over \sqrt{2}}}\big)
            d\varphi^2 + \sin^2\big({\textstyle{\chi\over \sqrt{2}}}\big)  \sin^2(2\varphi) \,
            d\psi^2 \,
}
with $\sqrt{g}=\sqrt{2} \sin^2({\chi\over \sqrt{2}})  \sin(2\varphi)\,
R^3$ with the radius $R=\sqrt k$. Thus, the semi-classical large
radius limit corresponds to $k\to\infty$.

Computing the total volume of the $S^3$ and comparing to other 
parametrizations of the $SU(2)$ we can fix the ranges of the variables
as
$0\le \chi \le \sqrt 2\, \pi$, $0\le \varphi \le  {\pi\over 2}$ 
and $0\le \psi \le  {2\pi}$.
Indeed we get $\int \sqrt{G}\, d\chi d\varphi d\psi=2\pi^2 R^3$.
Next we evaluate \eqref{hfluxtot} and obtain
\eq{
\label{SU(2)hflux}
   H=-k \sqrt 8 \sin^2\big({\textstyle{\chi\over \sqrt 2}}\big) \sin(2\varphi)\,
   d\chi\wedge d\varphi\wedge d\psi=-{\textstyle {2\over \sqrt k}} {\rm vol}(S^3)\,
}
so that the flux integral ${1\over (2\pi)^2} \int_{S^3} H =-k$ is
indeed quantized and the $H$-flux goes to zero for large $k$.
 
For the holomorphic one-forms \eqref{oneforms} on the $D$-brane one obtains
\eq{
    \theta^\alpha&=e^{i\psi} d\varphi +{i\over 2}\sin(2\varphi)\, e^{i\psi}   d\psi\,,\qquad
   \theta^{\ov\alpha}=e^{-i\psi} d\varphi -{i\over 2}\sin(2\varphi)\,
   e^{-i\psi} d\psi\,.
}
Inserting this into \eqref{metricbrane}, we indeed find  the metric
\eqref{SU2metric} on $S^3$ restricted to the $D$-brane world-volume
\eq{
          ds^2\vert_{\rm brane}=k\,\sin^2\big({\textstyle{\chi\over
             \sqrt{2}}}\big) \Big(4 d\varphi^2 +   \sin^2(2\varphi) 
            \, d\psi^2\Big) \,.
}
This metric describes an $S^2$ of radius $r=R\, \sin ({\chi\over \sqrt{2}})$. 
As described, for the $B$-field one can choose a gauge so that it has
only legs on the $D$-brane.  For $SU(2)$ this simply reads
\eq{
\label{SU2bfield}
B=k\,\Big( -\sqrt 2 \chi +
\sin\big( \sqrt{2} \chi\big) \Big)\sin(2\varphi)\, d\varphi\wedge d\psi
}
and via $H=dB$ indeed gives the $H$-flux from \eqref{SU(2)hflux}.
Clearly, the restriction of the $H$-flux to the brane is vanishing and
one can also show that 
the restriction of the $B$-field \eqref{SU2bfield} to the $D$-branes does not satisfy
the open string equation of motion. 
However, this latter point  can be reconciled  by also turning on a non-trivial gauge
flux on the brane
\eq{
F=dA={\sqrt 2 k\over 2\pi} \, \chi  \, \sin(2\varphi)\, d\varphi\wedge d\psi\,.
}
The gauge flux quantization condition ${1\over 2\pi} \int F\in \mathbb
Z$ leads to $\chi={2\pi\over k}{m\over \sqrt{2}}$ with $0\le  m\le k$,
which agrees with the formula below eq.\eqref{gaugeflux} by observing that the weight lattice of
$SU(2)$ is $\lambda=\mathbb Z/\sqrt{2}$ and that its Weyl-vector reads
$\rho=1/\sqrt{2}$. 
For the two
choices $m=0,k$, the co-dimension one $D$-brane degenerates to a
point-like $D$-brane  sitting at 
the north- or south-pole of the $S^3$, respectively.

Now, one can simply proceed by computing the globally defined two-form
flux ${\cal F}=B+2\pi F$ on the brane and the open string measure as
\eq{ 
\label{SU2measure}
{\cal F}=k\,\sin( \sqrt{2} \chi)  \sin(2\varphi) \, d\varphi\wedge
d\psi\,,\qquad
 \sqrt{ g+{\cal F} }=2 k \sin(2\varphi) \sin\big({\textstyle {\chi\over \sqrt
     2}}\big) \,.
}
Using  the dual holomorphic one-vectors 
\eq{
    \hat\theta_\alpha&={1\over 2} e^{-i\psi} \partial_\varphi -{i\over
      \sin(2\varphi)}\, e^{-i\psi}  \,\partial_\psi\,,\qquad
   \hat\theta_{\ov\alpha}={1\over 2} e^{i\psi} \partial_\varphi 
 + {i\over \sin(2\varphi)}\, e^{i\psi}  \,\partial_\psi 
}
and evaluating \eqref{bivector},  we get for the antisymmetric bi-vector $\Theta^{ij}$
on the brane
\eq{
\label{SU2bivector}
   \Theta=-{k\over 2 \sin(2\varphi)} \cot\left({\chi\over \sqrt
     2}\right) \, \partial_\varphi\wedge \partial_\psi\,.
}
Multiplying this with the measure \eqref{SU2measure} one realizes that
the $\varphi$ dependence drops out so that the open string equation of
motion \eqref{openeom}
is trivially satisfied on the $D$-brane. 

Since the brane is two-dimensional, one trivially has  $\Pi^{ijk}=0$, as well as
$\nabla_k \Theta^{ij}=0$. In appendix \ref{app_B} we also work out the 
$SU(3)$ WZW case and show that there one has a co-dimension two  $D$-brane supporting
a non-vanishing $\Pi^{ijk}$. Therefore, not all brane solutions of the
leading order string equations of motion necessarily have
$\Pi^{ijk}=0$ so that in our  later analysis we will also admit a
non-vanishing $\Pi^{ijk}$.

In the semi-classical limit, the two-dimensional world-volume of
the $D$-brane is expected to support a
non-commutative (but still associative) gauge theory. 
Since the world-volume is compact, for
fixed but large $k$, 
there can only be a finite number of quantum cells so that the
non-commutative gauge theory turned out not to be a field theory but 
rather  a matrix theory.
Using the operator product expansion of the corresponding vertex
operators, this theory has been derived in \cite{Alekseev:2000fd} and, as we 
discuss next, provides the first non-trivial application  of our L$_\infty$
bootstrap program.

 \subsection{L$_\infty$ algebra for the fuzzy 2-sphere}

Let us first briefly review some relevant features
of this  construction of the NC gauge theory in the fuzzy sphere limit
(for a little review see \cite{Alekseev:2000wg})
\eq{
\alpha' \to 0, \quad \alpha' k \to \infty \, .
}
This means that one takes  the zero-slope and the  large radius limit.

The rational boundary states are known explicitly and the open string
excitations at lowest energy are given in terms of the ground states
in the open string sector.  As discussed above, branes wrapping the
conjugacy classes $S^2\subset S^3$ are labelled by an integer $0\leq m
\leq k$. This integer determines a representation of the $SU(2)$
current algebra. In this section, we use instead the half-integer
representation labels $j$. The open string spectrum can  be organized
into the $SU(2)_k$ representations that appear in the fusion product of
$m$ with 
itself
\eq{
 \label{eq:su2decomp}
(j) \otimes (j) = \oplus_{j'=0}^{2j}\, (j') \, .
}
Here, one was working in  the large $k$ limit,  such that no
truncation appears in the fusion rules. It can now be observed  that
the representation $(j')$ contains $2j'+1$ ground states
$|Y_n^{j'}\rangle $ labeled by $j',n$ with  $n\in {-j', \dots , j'}$.
Geometrically, one can think about them in terms of spherical
harmonics on $S^2$. These finitely many states are  analogous 
to the infinitely many states $\exp(ik X)$ in the flat Moyal-Weyl case.

Note that the space of ground states is $(2j+1)^2$
dimensional. As proposed in \cite{Alekseev:1999bs}, these ground states  can
be identified with square matrices ${\rm Mat}(2j+1)$. $SU(2)$ acts on
these  matrices in the adjoint representation. This representation is not
irreducible; decomposing it into irreducible representations
reproduces precisely the  decomposition (\ref{eq:su2decomp}).

Furthermore, there is a product structure on the space of ground
states arising from the (truncated) operator product expansion (OPE) between the corresponding
vertex operators. Since their conformal weight $h_j=j(j+1)/(k+2)$ goes
to zero in the $k\to \infty$ limit,  the boundary OPE  becomes regular
in this limit.  
As it turns out \cite{Alekseev:2000fd}, the
information about these OPEs is precisely encoded in the
non-commutative matrix
product $f\cdot g$ in ${\rm Mat}(2j+1)$. This in particular allows to compute the
correlation functions of arbitrary vertex operators in terms of traces
over products of matrices. The upshot is that one deals with
an associative matrix algebra. The matrix product plays the role
of the Moyal-Weyl  star-product, so that we define
\eq{
              [f,g]:= f\cdot g- g\cdot f\,.
}
The above structure includes an action of angular momentum on the
spherical harmonics. It is obtained from the OPE  between the WZW
currents and  the vertex operators corresponding to the ground
states. From this one obtains for any matrix $A\in {\rm Mat}(2j+1)$
\eq{
L_a A = \frac{1}{\sqrt{2}} [Y_a^1 , A] \quad\,, a\in \{1,2,3 \} \, .
}
In the flat space limit, the operators $L_a$ can be thought of as
derivatives $L^a \to -i \partial^a$. However, these operators do not
commute but satisfy
\eq{
\label{fuzzycomm}
          [L_a,L_b]=i f_{ab}{}^c \, L_c\,
}
where $ f_{abc}$ are the totally antisymmetric $SU(2)$ structure constants. 

The effective theory on $N$ branes of type $j$ wrapping  this fuzzy
2-sphere was described as a gauge theory
with the gauge potential $A_a \in {\rm Mat}(N) \otimes {\rm Mat}
(2j+1)$. Here ${\rm Mat}(N)$ labels the Chan-Paton factors. 
This gauge field has to satisfy the physical state condition
$L^a A_a=0$.

Using CFT techniques, the effective action was shown to be a sum 
of a Yang-Mills term and a Chern-Simons term.
The two terms are separately  invariant under the following
gauge transformation
\eq{
\label{fuzzyvari}
\delta_f A_a = i L_a f + i [A_a, f] \, , 
}
where $f$ is an arbitrary matrix in ${\rm Mat}(N) \otimes {\rm
  Mat}(2j+1)$. Note that the derivative operator $L_a$ only acts
non-trivially on the degrees of freedom in ${\rm Mat}(2j+1)$. 
The closure of two such gauge variations gives
\eq{
\label{fuzzyclose}
              [\delta_f,\delta_g]=\delta_{i[f,g]}\,.
}
Introducing  a field strength
\eq{
F_{ab} = i L_a A_b - i L_b A_a + i [ A_a, A_b] + f_{ab}{}^c A_c \,,
}
the action can be expressed as 
\eq{
            S= {1\over 4} {\rm tr} \Big( F_{ab}\, F^{ab} \Big)-{i\over 2}
            {\rm tr} \Big( f^{abc}\, {\rm CS}_{abc} \Big)
}
with 
\eq{  
               {\rm CS}_{abc}=L_a A_b\, A_c +{1\over 3} A_a [A_b,A_c]
             -{i\over 2} f_{ab}{}^d A_d A_c\,.
}
The resulting equation of motion can  be written as
\eq{
\label{fuzzyeom}
0&=L^b F_{ba} + [A^b, F_{ba} ]  \\[0.1cm]
&= iL^b L_b A_a - i L^b (L_a A_b) - f_{a}{}^{bc} L_b A_c\\
&\phantom{=}+ iL^b [A_b, A_a] + [A^b, iL_b A_a-iL_a A_b ] - f_{a}{}^{bc} [A_b, A_c] \\ 
&\phantom{=} + i[A^b,[A_b,A_a]] \,.
}

As a first application of our approach, we now show that  the form of this  NC gauge theory on the fuzzy
sphere can be bootstrapped by invoking an L$_\infty$ structure.
The computation turns out to be  similar to the Moyal-Weyl
case  but includes  some corrections terms that can be traced back to the non-trivial 
commutator \eqref{fuzzycomm} of the derivatives.
Let emphasize that we proceed not just by simply checking the L$_\infty$
algebra  but by bootstrapping the higher products via
the  L$_\infty$ relations. Of course, one needs some initial
information to get started. 

As usual, we consider the graded vector space $X_0\oplus X_{-1} \oplus
X_{-2}$ with now matrix valued gauge parameters in $X_0$, gauge fields in
$X_{-1}$ and equations of motion in $X_{-2}$. 
From the gauge variation \eqref{fuzzyvari} and the closure condition
\eqref{fuzzyclose} we read-off
\eq{
\ell_1 (f) =i  L_a f , \qquad \ell_2 (f, g) = -i[f, g] \, ,
}
Then, imposing the  L$_\infty$ relation ${\cal J}_2(f,g)=0$ fixes
\eq{  
     \ell_2 (f, A) = i[A_a, f]  \,.
}
From the linear term in the equation of motion \eqref{fuzzyeom} we read-off
\eq{
\ell_1 (A) &=  \ell^{\rm YM}_1 (A) +\ell^{\rm CS}_1 (A) \\
 &= iL^b L_b A_a - i L^b (L_a A_b) - f_{a}{}^{bc} L_b A_c
}
where, as indicated, the first two terms come from the variation of the YM action
and the last term from the variation of the CS action.
First, after using \eqref{fuzzycomm} we realize that 
\eq{
                   \ell^{\rm YM}_1(\ell_1(f))=-L_a f \,,\qquad
                   \ell^{\rm CS}_1(\ell_1(f))=L_a f \,
}
so that only the combination of the two kinetic terms satisfies the 
relation ${\cal J}_1(f)=\ell_1(\ell_1(f))=0$. Therefore, if one 
were missing the contribution to the kinetic energy from the CS-term, one would
be forced to introduce it by the nilpotency condition in the L$_\infty$ algebra.
Moreover, one can
further simplify 
\eq{
\ell_1 (A) = iL^b L_b A_a - i L_a (L^b A_b)\,
}
where the second term actually vanishes by the physical state
condition $L^b A_b=0$. Next, we consider the L$_\infty$ relation
${\cal J}_2(f,A)=0$. A straightforward computation reveals that this
can be satisfied by defining
\eq{
       \ell_2(f,E)=i[E_a,f]
}
and 
\eq{
\label{koelle}
         \ell_2(A,B)&=-i L^b [A_b,B_a] -i [A^b,L_b B_a-L_a
         B_b] \\
    &\phantom{=}+ f_a{}^{bc} [A_b,B_c] +(A\leftrightarrow B)\,.
}
This looks very similar to the Moyal-Weyl case, except for the 
term in the second line.
Note that $-{1\over 2} \ell_2(A,A)$ gives precisely the order
$O(A^2)$ terms in the equation of motion \eqref{fuzzyeom}, that
we bootstrapped from an L$_\infty$ relation.

Next, we observe that the $\ell_2\ell_2$-terms in relations ${\cal J}_3(f,g,h)={\cal
  J}_3(f,g,A)={\cal J}_3(f,g,E)=0$ involve only matrix commutators so
that they can directly be satisfied by setting
\eq{
            \ell_3(f,g,A)=\ell_3(f,g,E)=\ell_3(f,A,B)=\ell_3(f,A,E)=0\,.
}
The only non-trivial relation is ${\cal J}_3(f,A,B)=0$. However, one
can check that the extra terms coming  from the second line in \eqref{koelle}
cancel against each other so that the computation is analogous
to the Moyal-Weyl case presented in section \ref{sec_CSYML}. 
Thus, this relation fixes
\eq{
              \ell_3(A,B,C)=&-i [A^b,[B_b,C_a] ] -i [B^b,[C_b,A_a] ]
                                   -i [C^b,[A_b,B_a] ]\\
                                   &-i [A^b,[C_b,B_a] ] -i [C^b,[B_b,A_a] ]
                                   -i [B^b,[A_b,C_a] ]
}  
which is again consistent with the order $O(A^3)$ term in the equation
of motion. From the  higher order relations only ${\cal
  J}_4(f,A,B,C)=0$ is not trivially satisfied, but eventually vanishes
by the Jacobi identity of the matrix commutator. 

Thus, after taking the initial data $\ell_1(f)$, $\ell_1(A)$ and
$\ell_2(f,g)$ we have bootstrapped the remaining terms appearing
in the gauge variations and the equations of motion by imposing
the relations of an L$_\infty$ algebra. In particular, we  found the extra correction
terms $(\sim\!\! f_{abc})$ in the equations of motion. 
We consider this as first compelling evidence that the form of  NC
gauge theories (arising from string theory) is governed by 
an L$_\infty$ structure.
We will continue to elaborate on this idea in section \ref{sec_main}.

\subsection{Non-constant $\Theta$ via integrable deformations}

As we have seen, branes in  WZW models 
can lead to a non-constant $\Theta$ on a curved space. However,
for  compact group manifolds the effective NC-gauge theory in the
large volume limit is  rather a matrix model than a NC field theory
based on a star-product with non-constant $\Theta$.

In this section, we recall that recently string theory examples have
been identified that  are supposed to give rise to NC-field
theories on flat Minkowski space with non-constant $\Theta$.
These appeared in the context of integrable deformations
of the AdS$_5$ sigma model \cite{Bena:2003wd}. Here, we do not want
to review the whole construction and its refinements, as we are only interested in 
one of its aspects.

It was shown that from certain solutions to the classical Yang-Baxter
equation one can extract  a closed string metric $g_{ij}$, Kalb-Ramond
field $B_{ij}$ and dilaton $\phi$ (and R-R fields) that satisfies the string equations
of motion and, for the deformation going to zero, gives back the AdS$_5$ 
geometry. This construction can be seen as  a generalization of the early
analysis \cite{Hashimoto:1999ut,Maldacena:1999mh} of the supergravity
background dual to the Moyal-Weyl NC-gauge theory. 

Expressing the deformed solution in the open string frame
\cite{vanTongeren:2015uha, vanTongeren:2016eeb, Araujo:2017jkb,Araujo:2017jap}
revealed that the open string metric is still the one on AdS$_5$, the open
string dilaton is constant and the only change is the anti-symmetric bi-vector 
$\Theta$. There are cases, where the latter restricted to the boundary 
$\mathbb R_{1.3}\subset {\rm AdS}_5$ satisfies indeed the open
string equation of motion \eqref{openeom}. Therefore, one expects that
the gravity theory in the bulk is dual to a NC-gauge theory on the
flat boundary with non-constant $\Theta^{ij}$.

Here, we just present one example that we took from \cite{Araujo:2017jkb,Araujo:2017jap}. Taking
one specific  solution 
to the classical Yang-Baxter equation, in the open string frame gives
a flat metric with 
\eq{
           \Theta(x)=\left(\begin{matrix}  0 & 0 & -\eta x_1 x_3 & \eta
               x_1 x_2 \cr
                0 & 0 & -\eta x_0 x_3 & \eta x_0 x_2 \cr
                    \eta x_1 x_3 & \eta x_0 x_3 & 0 & 0  \cr
                   -\eta x_1 x_2 & -\eta x_0 x_2 & 0 & 0  \cr  
                    \end{matrix}\right)\,.
}
Here $\eta$ is the deformation parameter. One can readily check
that $\partial_i \Theta^{ij}=0$ is satisfied for all $j=0,\ldots,3$
and that $\Pi^{ijk}=0$.
Therefore, string theory admits solutions
that are expected to give rise to  NC gauge theories on flat (or
curved) spaces. Here we just cite one comment from \cite{vanTongeren:2015uha}:
'While  likely  technically
involved, we believe it should in principle be possible to construct (supersymmetric) gauge
theories on such non-commutative spaces using the methods 
developed in \cite{Seiberg:1999vs,Madore:2000en,Jurco:2001rq}'.
Our proposal rather is that such theories can be bootstrapped via L$_\infty$ 
algebras.

\section{L$_\infty$ structures for non-constant flux}
\label{sec_main}

As we have just seen, the question now arises whether one can deform 
the Moyal-Weyl case and formulate a consistent  NC-gauge theory
for non-constant $\Theta$. 
Motivated also by the previous example of branes in the $SU(2)$ WZW
model, we investigate whether the L$_\infty$ bootstrapping approach works and
gives reasonable results.
In this main section of the paper, we investigate this novel approach for  flat/curved space
with a completely generic non-constant $\Theta^{ij}(x)$ with even
non-vanishing $\Pi^{ijk}$.

First we  consider just the action of
the gauge symmetry and its closure, i.e. we  construct
the corresponding L$_\infty^{\rm gauge}$ algebra. In appendix
\ref{app_B} we explicitly  show
that this can be refined to an  A$_\infty^{\rm gauge}$ algebra.
First, let us explain that  for non-constant flux a serious problem
appears, for which L$_\infty$ offers a new solution.

\subsection{An issue for non-constant $\Theta$}

In the previous computations in section \ref{sec_gaul} it was essential that the
non-commuta\-tivity structure $\Theta^{ij}$ was constant. 
For non-constant $\Theta^{ij}$, even at linear order one runs into
the following issue concerning the Leibniz-rule.
Consider the generic star-product between functions 
\eq{
     f\bullet g=f\cdot g +  {i\over 2}\Theta^{ij}(x)\, \partial_i f\, \partial_j g +
  O(\Theta^2)
}
and apply $\partial_a$. One finds
\eq{
\label{changeleibniz}
     \partial_a (f\bullet g) = \partial_a f \bullet g + f\bullet \partial_a
     g + {i\over 2}(\partial_a \Theta^{ij})\, \partial_i f \partial_j g + O(\Theta^2)\,,
}
i.e. the derivative does not satisfy the usual Leibniz-rule with
respect to the star-product. One way to resolve this issue is
to generalize the co-product, leading to the general structure of a
Hopf-algebra. For this purpose one recalls that for the usual product
of functions $\mu: {\cal A}\otimes {\cal A}\to {\cal A}$ 
with 
\eq{
              \mu(f\otimes g)=  f\cdot g
}
the enveloping algebra ${\cal H}$ of the variations $\delta_a=\partial_a$ defines a Hopf-algebra with
product $m:{\cal H}\otimes {\cal H}\to {\cal H}$ 
\eq{
             m(\delta_a\otimes \delta_b)= \delta_a \, \delta_b 
}
and co-product $\Delta:{\cal H}\to {\cal H}\otimes {\cal H}$
\eq{
                  \Delta(\delta_a)=\delta_a\otimes 1 + 1\otimes
                  \delta_a
}
where we have used the action ${\cal H}\otimes {\cal A}\to {\cal A}$
with $\delta_a(f)=\partial_a f$. For consistency, the co-product should be co-associative,
\eq{
\left(\Delta\otimes
  1\right)\circ\Delta=\left(1\otimes\Delta\right)\circ\Delta\, , 
\label{coassoc}
}
as well as admit a co-unit and an antipode, see e.g. \cite{Szabo:2006wx}  for details.
Then, the Leibniz-rule can be abstractly
written as
\eq{
                 \delta_a (\mu(f\otimes g) )=\mu\circ
                 \Delta(\delta_a) (f\otimes g)\,.
}
The relation \eqref{changeleibniz} can be written in an analogous manner by
defining the new product
\eq{
              \mu^\star(f\otimes g)=  f\bullet g
}
between functions and the adjusted or deformed co-product $\Delta^\star(\delta_a)$
for elements in ${\cal H}$. Then \eqref{changeleibniz} can still be
written as
\eq{
                 \delta_a (\mu^\star(f\otimes g) )=\mu^\star\circ
                 \Delta^\star(\delta_a) (f\otimes g)\,.\label{deformed-Leibnitz}
}
Note, that the consistent definition of the deformed co-product
satisfying the co-associativity condition \eqref{coassoc} and the
relation \eqref{deformed-Leibnitz} is a highly nontrivial problem. The
known solution is to use an  invertible element $\mathscr{F}\in {\cal
  H}\otimes {\cal H}$, called a twist, with the help of which the
original star product can be  represented as 
\eq{f\bullet g= \mu^\star(f\otimes g)=\mu \circ{\mathscr{F}}^{-1}(f\otimes g)\ .}
Then the deformed co-product is given by 
\eq{ \Delta^\star(\delta_a)={\mathscr{F}}\Delta(\delta_a){\mathscr{F}}^{-1}\ .}
However, only  very few examples of star products originating from a  twist are known.

Another point which should be mentioned here is that the deformed
co-product is still co-associative and that is why in the Hopf-algebra
approach no higher products or brackets are needed to compensate the
violation of the original Leibniz rule. This is the key difference
with our proposal in this paper. Nevertheless, we leave for the future
a  better understanding of the relation between our approach and
other previous approaches to the construction of non-commutative gauge 
theories.

In view of the proposal that generic (gauge) symmetries in string
theory are related to L$_\infty$ structures, let us have a second look 
at the violation of the naive Leibniz-rule \eqref{changeleibniz}. 
Recall that for NC-Yang-Mills theory we found $\ell_1(f)=\partial_a
f$. If we define $\ell_2(f,g)=i[f,g]_\bullet=i(f\bullet g- g\bullet f)\in X_0$ then by
anti-symmetrization the relation  \eqref{changeleibniz}
is closely related to 
\eq{
\label{changeleibnizb}
    \ell_1( \ell_2 (f,g))&= i[\overbrace{\ell_1(f)}^{\in X_{-1}}, g]_\bullet + i[f, \overbrace{\ell_1(g)}^{\in X_{-1}}]_\bullet
    - (\partial_a \Theta^{ij})\, \partial_i f \partial_j g +
    O(\Theta^2)\,,\\[0.2cm]
       &= \ell_2(\ell_1(f),g)+\ell_2(f,\ell_1(g))\,.
}
From this point of view, the correction term $\partial \Theta$ only
indicates that we should better not define
$\ell_2(f,A)=i[f,A]_\bullet\in X_{-1}$ but instead
\eq{
            \ell_2(f,A)=i[f,A_a]_\bullet -{1\over 2} (\partial_a \Theta^{ij})\, \partial_i f A_j  +O(\Theta^2)\,.
}
Note that in the L$_\infty$ algebra, $\ell_2(f,g)\in X_0$ and $\ell_2(f,A)\in X_{-1}$
are a priori {\it different} products. 
Thus, we can still satisfy the usual L$_\infty$ Leibniz-rule by
changing the action of the NC-gauge symmetry on the gauge fields
\eq{
     \delta_f A&=\ell_1(f) +\ell_2(f,A) + \ldots\\
     &= \partial_a f + i[f,A_a]_\bullet -{1\over 2} (\partial_a
     \Theta^{ij})\, \partial_i f A_j + \ldots\,.
}
By construction, this guarantees that the closure condition
\eq{
          [\partial_f,\partial_g] A= \delta_{-i[f,g]_\bullet} A
}
is indeed satisfied up to linear order in $\Theta$.

Proceeding in this way, the higher products of the L$_\infty$ algebra 
will receive higher derivative corrections, leading to corrections to the
action of the symmetries on the gauge fields and eventually to the equations of motion.
This latter approach seems to be completely  different from the resolution of the problem
via twisted symmetries and Hopf-algebras and much closer to the
structure of symmetries in string theory.
In the remainder of this section, we will work this out  in more detail
and compute the corresponding derivative corrections to
L$_\infty^{\rm gauge}$   up to second order in $\Theta$.
At this order also the non-associativity enters.

\subsection{L$_{\infty}^{\rm gauge}$ algebra at order $O(\Theta^2)$ }
\label{sec_Lgauge}

The resolution of the Leibniz-rule was done at linear order in $\Theta$
and it is of course not clear whether the procedure can be
consistently continued  to higher orders in $\Theta$. To get some
confidence, starting with the Kontsevich star-product \eqref{starkont}
at second order in $\Theta$, 
in this section we construct the corresponding L$_{\infty}^{\rm gauge}$ algebra.

Thus, we have  the vector space $X_0\oplus X_{-1}$ containing gauge
parameters and gauge fields still choose
\eq{
          \ell_1(f)=\partial_a f\,.
}
Moreover, for $\ell_2(f,g)$ we want to have the Kontsevich
star-commutator
\eq{
         \ell_2(f,g)=i(f\bullet g- g\bullet f)=- \Theta^{ij} \partial_i
         f \partial_j g+O(\Theta^3)\,.
}
Note that the even order terms in $\Theta$ drop out in the
star-commutator. Therefore, the analysis of the Leibniz rule ${\cal
  J}_2(f,g)=0$ from  the previous section is still valid and we get
\eq{
             \ell_2(f,A)=
            i[f,A_a]_\bullet -{1\over 2} (\partial_a \Theta^{ij})\, \partial_i f A_j  +O(\Theta^3)\,.
}
Next, we have to impose the L$_\infty$ relations ${\cal J}_3(f,g,h)=0$
and  ${\cal J}_3(f,g,A)=0$.
The first relation explicitly reads
\eq{
\label{linfrelay}
0&=\ell_2(\ell_2(f,g),h)+\ell_2(\ell_2(g,h),f)+\ell_2(\ell_2(h,f),g)+\\[0.1cm]
&\phantom{=i}\ell_3(\ell_1(f),g,h)+\ell_3(f,\ell_1(g),h)+\ell_3(f,g,\ell_1(h))\ .
}
The first line is just the Jacobiator for the star-commutator and
yields
\eq{
          -\Pi^{ijk} \partial_i f\partial_j g\partial_k h
}
that we do not require to be vanishing.
Taking into account the graded symmetry of the brackets we find 
\begin{equation}
\ell_3(A,f,g)=\frac13\Pi^{ijk} A_i\partial_jf\partial_kg\ ,
\end{equation}
Next, we have to analyze the relation  ${\cal J}_3(f,g,A)=0$, which is
explicitly given by
\eq{
0&=\ell_2(\ell_2(A,f),g)+\ell_2(\ell_2(f,g),A)+\ell_2(\ell_2(g,A),f)+\\[0.1cm]
&\phantom{=i}\ell_1(\ell_3(A,f,g))-\ell_3(A,\ell_1(f),g)-\ell_3(A,f,\ell_1(g))\ 
}
where we used $\ell_1(A)=0$. The last two terms involve the
three-product $\ell_3(A,B,f)$ that needs to be determined from this relation.
For this purpose, we  calculate
\eq{ 
\label{halfrela}
&\ell_2(\ell_2(A,f),g)+\ell_2(\ell_2(f,g),A)+\ell_2(\ell_2(g,A),f)+\ell_1(\ell_3(A,f,g))
  =\\[0.1cm]
&=-{1\over 2} G_a{}^{ijk}\, A_i \,\partial_j
  f \,\partial_k g  -\Pi^{ijk}\, \partial_i A_a \,\partial_j f \,\partial_k g 
+{1\over 3}  \Pi^{ijk} \,\partial_a A_i\, \partial_j f\, \partial_k g \\[0.1cm]
&\phantom{=i} +{1\over 3}  \Pi^{ijk}\, A_i \,\partial_a\partial_j f\, \partial_k g 
   +{1\over 3}  \Pi^{ijk} A_i  \,\partial_j f \,\partial_a\partial_k g
}
 with
\eq{
G_a{}^{ijk}= {1\over 3}\partial_a \Pi^{ijk} -
  \Theta^{im} \partial_m \partial_a \Theta^{jk} -{1\over
    2} \partial_a\Theta^{jm} \partial_m\Theta^{ki}-{1\over
    2} \partial_a\Theta^{km} \partial_m\Theta^{ij}
}
that apparently satisfies the relation
\eq{
\label{nicerelaG}
        G_a{}^{ijk}+G_a{}^{kij}+G_a{}^{jki}=0\,.
}
Taking into account the symmetry $\ell_3(A,B,f)=\ell_3(B,A,f)$, \eqref{halfrela}
motivates to make the ansatz
\eq{
   \ell_3(A,B,f)&=\alpha (G_a{}^{ijk} + G_a{}^{jik})  A_i
   B_j \partial_k f \\[0.1cm]
   &+\beta \Pi^{ijk} (\partial_a A_i B_j  \partial_k f   -
   A_i \partial_a B_j  \partial_k f   )\\[0.1cm]
&+\gamma \Pi^{ijk} (\partial_i A_a B_j  \partial_k f -
  A_i \partial_j  B_a  \partial_k f)\,.
}
Inserting this into the L$_\infty$ relation \eqref{linfrelay}, using \eqref{nicerelaG}
and comparing coefficients we find
\eq{
        \alpha=-{1\over 6}\,,\qquad \beta={1\over 6}\,,\qquad \gamma=-{1\over 2}
\,.}
Note that for $\beta$ and $\gamma$ we have 3 different equations for
the two unknowns. Moreover, the relation \eqref{nicerelaG} 
was used to solve for $\alpha$, and is in fact the consistency condition for the existence of the solution of the equation \eqref{linfrelay}, see \cite{Kupriyanov:2008dn} for more details. Thus, it is highly
non-trivial that indeed the L$_\infty$ relation 
${\cal J}_3(f,g,A)=0$ can be satisfied via
\eq{
     \ell_3(A,B,f)&=-{1\over 6}\Big(G_a{}^{ijk}+G_a{}^{jik}\Big) A_i
     B_j \partial_k f\\[0.1cm]
    &+{1\over 6} \Pi^{ijk}(\partial_a A_i B_j \partial_k f - 
    A_i \partial_a B_j \partial_k f )\\[0.1cm]
    &-{1\over 2} \Pi^{ijk}(\partial_i A_a B_j \partial_k f - 
    A_i \partial_j B_a \partial_k f )\,.
}      
Note that, as opposed to  $\ell_3(A,f,g)$, the three-product  $ \ell_3(A,B,f)$ is non-vanishing
in the associative case, either. 
Recall that our computation was exact only up to second order in
$\Theta$. Setting now all higher products to zero up to this order, 
all higher relations  ${\cal J}_n=0$ for $n\ge 4$ are
automatically satisfied, as well. This is because $\ell_2 \ell_3$ is already 
third order in $\Theta$.  
Therefore, for non-constant $\Theta$ we have constructed a consistent
L$_\infty^{\rm gauge}$ algebra, for which derivative $\partial\Theta$
corrections induce non-vanishing higher products (even in the
associative case). This is very compelling, as in the course of this
computation there arose  non-trivial consistency conditions that just happened to
be satisfied. 

We also analyzed whether the gauge  structure features an underlying 
A$_\infty$ algebra. Since the higher products are not any longer
graded symmetric,   for that purpose one has to determine many more individual higher
products that are also constrained  by  more A$_\infty$ relations. In this sense, an
A$_\infty$ structure
can be considered as a refinement of an L$_\infty$ structure.
Since the computations turned out to be quite
lengthy and involved, we delegated the presentation of the positive results into appendix
\ref{app_B}.

Let us proceed and  extend the former L$^{\rm gauge}_\infty$  algebra for the action of
a gauge symmetry on gauge fields by a vector space $X_{-2}$ that
contains the equation of motion of a star-deformed 3D Chern-Simons
theory  and a star-deformed $U(1)$ Yang-Mills theory.

\subsection{NC Chern-Simons theory}

Let us first consider a NC-CS theory. Since a metric does not appear
neither in the Kontsevich star-product nor in the topological
CS-theory, we suspect that the following considerations
are valid irrespective of a  metric.
Let us first consider the appearing structure 
up to linear order in $\Theta$. From the former
discussion, we expect to find  also
here $\partial \Theta$ corrections.

As said, we have three vector spaces like
\eq{      
\begin{matrix}   X_0\quad  &\quad X_{-1}\quad  &\quad  X_{-2} \\[0.2cm]
                        f    \quad  &\quad   A_a   \quad  &\quad  E_a
\end{matrix}
}
with the following derivatives 
\eq{
           \ell_1(f) = \partial_a f\,,\qquad
           \ell_1(A)=\epsilon_c{}^{ab} \, \partial_a A_b 
}
that clearly satisfy $\ell_1(\ell_1(f))=0$. With 
\eq{
             \ell_2(f,g)=i[f,g]_\bullet=-\Theta^{ij} \partial_i f \partial_j g
             +O(\Theta^2)
}
we have already seen that the Leibniz-rule
$\ell_1(\ell_2(f,g))=\ldots$ fixes 
\eq{
        \ell_2(f,A)=i[f,A_a]_\bullet-{1\over 2}
        \partial_a \Theta^{ij} \partial_i f A_j+O(\Theta^2)\,.
}
Next, one has to check the Leibniz rule 
\eq{
    \ell_1(\ell_2(f,A))=\ell_2(\ell_1(f),A)+\ell_2(f,\ell_1(A)) \,.
}
By making an ansatz for $\ell_2(A,B)$ and $\ell_2(f,E)$, using
that the former is symmetric  and fixing the
coefficients one finds 
\eq{
       \ell_2(f,E)=i[f,E_a]_\bullet+O(\Theta^2)
}
and
\eq{
        \ell_2(A,B)=\;\epsilon_c{}^{ab}\,  i[A_a,B_b]_\bullet  
    &-\epsilon_c{}^{ab} \partial_a \Theta^{ij} \Big( A_i \partial_j
    B_b + B_i \partial_j A_b \Big)\\
    &+{\textstyle {1\over 2}} \epsilon_c{}^{ab} \partial_a \Theta^{ij} \Big( A_i \partial_b
    B_j + B_i \partial_b A_j \Big)+O(\Theta^2)\,.
}
Note that indeed $\ell_2(A,B)$ is symmetric under exchange of the two
arguments and that $\ell_2(f,E)$ does not receive  any
$\partial\Theta$ correction.
The Leibniz-rules for $(AA), (fE), (AE)$ are trivially satisfied, as
they lie in trivial vector spaces $X_{-3}$ and  $X_{-4}$.
Since all $\ell_2$ products are linear in $\Theta$, all $\ell_2\circ
\ell_2$ relations are trivially satisfied up to linear order in $\Theta$.

Therefore, by requiring the consistency of an underlying L$_\infty$
algebra for the case of a $\Theta$ deformed 3D Chern-Simons theory,
we have extracted a $\partial\Theta$ correction to the equation of
motion 
\eq{
           {\mathpzc F}&=\ell_1(A) -{1\over 2} \ell_2(A,A) +O(\Theta^2)\\
     &=\epsilon_c{}^{ab} \bigg( \partial_a A_b -{\textstyle {i\over 2}} [A_a,A_b]_\bullet
    +\partial_a \Theta^{ij}  \Big(A_i \partial_j A_b 
    -{\textstyle {1\over 2}} A_i \partial_b A_j \Big)\bigg)+O(\Theta^2)\,.
}
We find it tantalizing that the algebraic structure of an L$_\infty$
algebra allowed
one to fix derivative corrections to the equations of motion
of a non-commutative CS theory. 
Of course, so far  this computation is only up to linear order in
$\Theta$,
but we conjecture  that it can be extended in a consistent way to higher
orders. As a non-trivial check let us now consider the corrections
at second order in $\Theta$. This is the first instance where 
also the associators appear.

\subsubsection*{NC-CS at order $O(\Theta^2)$}

The star commutator to this order remains unchanged, so do the
previously defined structures for $\ell_2$. 
However, at  this order there appear higher brackets $\ell_3$. The
expressions for $\ell_3(A, f,g)$ and $\ell_3(A,B,f)$ were found in
Sect. 3.1. Taking into account that now $X_{-2}$ is non trivial, one
may also have non-vanishing $\ell_3(E,f,g)\in X_{-1}$,
$\ell_3(E,A,f)\in X_{-2}$ and $\ell_3(A,B,C)\in X_{-2}$.

Let us start with $\ell_3(E,f,g)$. Such a term contributes to the
closure condition ${\cal J}_3(f,g,A)=0$, which are however satisfied 
without it. Therefore, we can set $\ell_3(E,f,g)=0$.
Next  we consider ${\cal J}_3(E,f,g)=0$ 
\eq{
&0=\ell_2(\ell_2(E,f),g)+\ell_2(\ell_2(g,E),f)+\ell_2(\ell_2(f,g),E)\\[0,1cm]
&\phantom{=i}+\ell_3(E,\ell_1(f),g)+\ell_3(E,f,\ell_1(g))
}
from which we derive
\eq{
\ell_3(E,A,f)=\frac12\  \Pi^{ijk}\partial_i E_a A_j \partial_k f\, .
}
It is understood, that all expressions are (only) correct up to order $O(\Theta^2)$.
Finally, to determine $\ell_3(A,B,C)$, we consider ${\cal J}(A,B,f)$
and write is as
\eq{
&\ell_3(A,B,\ell_1(f))=\\
&-\ell_1(\ell_3(A,B,f))-\ell_3(\ell_1(A),B,f)+\ell_3(A,\ell_1(B),f)\\
&-\ell_2(\ell_2(A,B),f)-\ell_2(\ell_2(f,A),B)+\ell_2(\ell_2(B,f),A)\,.
}
The right hand side of this relation is quite involved, so we
follow the same strategy as in the section \ref{sec_Lgauge} and  collect
structures with the same number of derivatives acting on the arguments
$A$, $B$ and $f$.  In principle we can get terms from one up to four
derivates so that we write
\eq{
\ell_3(A,B,\ell_1(f))=\sum_{N=1}^4 \ell^{(N)}_3(A,B,\ell_1(f))\,.
}
After a tedious computation we find a couple of cancellations and
simplifications. First we obtain that two of the four terms do vanish
\eq{
          \ell^{(1)}_3(A,B,\ell_1(f))=\ell^{(4)}_3(A,B,\ell_1(f))=0\,.
}
The term with two derivatives using the relation \eqref{nicerelaG} can be written in the convenient way
\eq{
        \ell^{(2)}_3(A,B,\ell_1(f))=      
   -\epsilon_c{}^{ab}&\Big( 
   {\textstyle {1\over 3}} \Theta^{km} \partial_b \partial_m \Theta^{ij}
   -{\textstyle {1\over 6}} \partial_b \Theta^{km} \partial_m \Theta^{ij} +(j
   \leftrightarrow k)\Big)\\
    &\Big(\partial_a A_i B_j \partial_k f  + A_j \partial_a
    B_i \partial_k f\Big) \\[0.1cm]
 -  \epsilon_c{}^{ab}&\Big( 
   {\textstyle -{1\over 2}} \Theta^{km} \partial_b \partial_m \Theta^{ij}
   +{\textstyle {1\over 2}} \partial_b \Theta^{km} \partial_m \Theta^{ij} +(j
   \leftrightarrow k)\Big)\\
    &\Big(\partial_i A_a B_j \partial_k f  + A_j \partial_i
    B_a \partial_k f\Big) \\[0.1cm]
   -\epsilon_c{}^{ab}&\Big( 
   -{\textstyle {1\over 6}} \Theta^{km} \partial_b \partial_m \Theta^{ij}
   +{\textstyle {1\over 3}} \partial_b \Theta^{km} \partial_m \Theta^{ij} +(j
   \leftrightarrow k)\Big)\\
    &\Big( A_j B_k \partial_a\partial_i f  \Big)\\[0.1cm]
 - \epsilon_c{}^{ab} &\Big( {\textstyle {1\over 2}}  \partial_a
  \Theta^{ij} \partial_b \Theta^{kl}\Big) \\
   & \Big( (\partial_i A_k -\partial_k A_i) B_j \partial_l f +
                     A_j (\partial_i B_k -\partial_k B_i) \partial_l f\Big)\,.
}
Note that this is explicitly symmetric under the exchange of the gauge
fields $A$ and $B$.
Moreover, the numerical prefactors are just right  to directly read off
a totally symmetric three product
\eqb{
\label{threeprodseca}
        \ell^{(2)}_3(A,B,C)=  
      -\epsilon_c{}^{ab}&\Big( 
   {\textstyle {1\over 3}} \Theta^{km} \partial_b \partial_m \Theta^{ij}
   -{\textstyle {1\over 6}} \partial_b \Theta^{km} \partial_m \Theta^{ij} +(j
   \leftrightarrow k)\Big)\\
    &\Big(\partial_a A_i B_j C_k  + A_j \partial_a
    B_i C_k   +A_k B_j \partial_a C_i  \Big) \\[0.1cm]
- \epsilon_c{}^{ab}&\Big( 
   {\textstyle -{1\over 2}} \Theta^{km} \partial_b \partial_m \Theta^{ij}
   +{\textstyle {1\over 2}} \partial_b \Theta^{km} \partial_m \Theta^{ij} +(j
   \leftrightarrow k)\Big)\\
    &\Big(\partial_i A_a B_j C_k + A_j \partial_i
    B_a C_k +A_k B_j \partial_i C_a \Big) \\[0.1cm]
  - \epsilon_c{}^{ab} &\Big( {\textstyle {1\over 2}}  \partial_a
  \Theta^{ij} \partial_b \Theta^{kl}\Big) \\
   & \Big( (\partial_i A_k -\partial_k A_i) B_j C_l+
                     A_j (\partial_i B_k -\partial_k B_i) C_l\\
    &\phantom{a}   +A_l B_j (\partial_i C_k -\partial_k C_i)
    \Big)\,.
 }
Next we come to the contribution with three derivatives. Here one has 
essentially two terms, one proportional to $\Pi^{ijk}$ and one that does
not vanish in the associative case
\eq{
   \ell^{(3)}_3(A,B,\ell_1(f))=      
   \epsilon_c{}^{ab} \Pi^{ijk}&\Big(\;
{\textstyle {1\over 3}} \partial_a
     A_i \partial_b B_j \partial_k f+ \partial_i
     A_a \partial_j B_b \partial_k f\\
 &+{\textstyle {1\over 6}} \big(\partial_a A_i  B_j \partial_b \partial_k
       f+ A_j  \partial_a B_i \partial_b \partial_k f\big)\\
    &    -{\textstyle {1\over 2}} \big(\partial_i A_a  B_j \partial_b \partial_k
       f+ A_j  \partial_i B_a \partial_b \partial_k f\big)\\
    &-{\textstyle {1\over 2}} \big(\partial_i A_a  \partial_b B_j  \partial_k
       f+ \partial_b A_j  \partial_i B_a \partial_b \partial_k f\big)
       \Big)\\[0.1cm]
 -\epsilon_c{}^{ab} \Theta^{kl}\partial_b \Theta^{ij}&\Big(\;
  {\textstyle {1\over 2}} \big( \partial_k A_a
     B_i \partial_j\partial_l f +   A_i \partial_k
     B_a \partial_j\partial_l f \big) \\
 &-{\textstyle {1\over 2}} \big( \partial_k A_a
    \partial_l  B_i \partial_j  f +  \partial_l  A_i \partial_k
     B_a \partial_j f \big) \\
 &-{\textstyle {1\over 2}} \big( \partial_l A_i
     B_j \partial_k\partial_a f +   A_j \partial_l
     B_i \partial_k\partial_a f \big) 
\Big)\,.
}
Again the relative coefficients are just right to define a totally
symmetric three-product
\eqb{
\label{threeprodsecb}
     \ell^{(3)}_3(A,B,C)= {\textstyle {1\over 3}} \epsilon_c{}^{ab} \Pi^{ijk}&\Big(\;
 \partial_a  A_i \partial_b B_j C_k + A_i \partial_a B_j \partial_b
 C_k +\partial_b A_i B_j \partial_a C_k\Big)\\
+ \epsilon_c{}^{ab} \Pi^{ijk}&\Big(\;
 \partial_i  A_a \partial_j B_b C_k + A_i \partial_j B_a \partial_k
 C_b +\partial_i A_b B_j \partial_k C_a\Big)\\
-{\textstyle{1\over 2}}\epsilon_c{}^{ab} \Pi^{ijk}&\Big(\;
 \partial_i  A_a \partial_b B_j C_k + A_i \partial_j B_a \partial_b
 C_k +\partial_b A_i B_j \partial_k C_a\\
&+
 \partial_a  A_i \partial_j B_b C_k + A_i \partial_a B_j \partial_k
 C_b +\partial_i A_b B_j \partial_a C_k\Big)\\
+ {\textstyle{1\over 2}}\epsilon_c{}^{ab} \Theta^{kl}\partial_b \Theta^{ij}&\Big(\;
 \partial_k  A_a (\partial_l B_i C_j - B_i \partial_l C_j)+
 \partial_k  B_a (\partial_l A_i C_j - A_i \partial_l C_j)\\
&+ \partial_k  C_a (\partial_l A_i B_j - A_i \partial_l B_j) \Big)\,.
}

\vspace{0.4cm}
\noindent
Therefore invoking the L$_\infty$ relations, in \eqref{threeprodseca} and \eqref{threeprodsecb}  we
have determined a  totally symmetric three-product 
$\ell_3(A,B,C)=\ell^{(2)}_3(A,B,C)+\ell^{(3)}_3(A,B,C)$.
Note that all the formerly determined higher products went into this
computation and that all the prefactors just came out right to fit
into the L$_\infty$ structure. This procedure works both for the
associative and the non-associative cases. Even in the associative case
the three-product is non-vanishing and receives derivative corrections.

Moreover, all higher L$_\infty$ relations are automatically satisfied
at $O(\Theta^2)$. This is all  very compelling and makes us believe
that this procedure can be continued to higher orders in $\Theta$.
As a consequence, up to quadratic order, the equation of motion of the NC-CS
gauge theory reads
\eq{
           {\mathpzc F}&=\ell_1(A) -{1\over 2} \ell_2(A^2) -{1\over 3!}
           \ell_3(A^3) +O(\Theta^3)\\
     &=\epsilon_c{}^{ab} \bigg[ \partial_a A_b -{\textstyle {i\over 2}} [A_a,A_b]_\bullet
    +\partial_a \Theta^{ij}  \Big(A_i \partial_j A_b 
    -{\textstyle {1\over 2}} A_i \partial_b A_j \Big)\\
     & \phantom{=\epsilon_c{}^{ab}} +{1\over 6}\Big( 
    \Theta^{km} \partial_b \partial_m \Theta^{ij}
   -{\textstyle {1\over 2}} \partial_b \Theta^{km} \partial_m \Theta^{ij} +(j
   \leftrightarrow k)\Big) \Big(\partial_a A_i A_j A_k \Big) +\\[0.1cm]
 &\phantom{=\epsilon_c{}^{ab}} -{1\over 4} \Big( 
    \Theta^{km} \partial_b \partial_m \Theta^{ij}
   - \partial_b \Theta^{km} \partial_m \Theta^{ij} +(j
   \leftrightarrow k)\Big)\Big(\partial_i A_a A_j A_k \Big)+ \\[0.1cm]
  &\phantom{=\epsilon_c{}^{ab}} + {1\over 2}\Big(  \partial_a
  \Theta^{ij} \partial_b \Theta^{kl}\Big)\Big( \partial_i A_k  A_j A_l
  \Big)\\
&\phantom{=\epsilon_c{}^{ab}} -{\textstyle {1\over 2}} \Pi^{ijk}\Big(\;
 {\textstyle {1\over 3}}\partial_a  A_i \partial_b A_j A_k 
+\partial_i  A_a (\partial_j A_b-\partial_b A_j)  A_k 
\Big)\\
&\phantom{=\epsilon_c{}^{ab}} - {\textstyle{1\over 2}}\Big(\Theta^{kl}\partial_b \Theta^{ij}\Big)\Big(\;
 \partial_k  A_a \partial_l A_i A_j  \Big) \bigg]+ O(\Theta^3)\,.
 }
 Let us emphasize that this is designed such that it transforms
covariantly under an $(\partial\Theta)$ corrected NC gauge transformations 
\eq{
 \delta_f A_a&=\ell_1(f)+\ell_2(f,A)-{1\over 2} \ell_3(f,A,A) + O(\Theta^3)\\
                   &=\partial_a f+i[f,A_a]_\bullet -{1\over
                     2} \partial_a \Theta^{ij} \partial_i f A_j\\
&\phantom{=} -{1\over 6}\Big(
\Theta^{im} \partial_a \partial_m \Theta^{jk}
   -{\textstyle {1\over 2}} \partial_a \Theta^{im} \partial_m \Theta^{jk} \Big) A_i
     A_j \partial_k f\\[0.1cm]
&\phantom{=} -{1\over 2} \Pi^{ijk}\Big({\textstyle {1\over 3}}\partial_a A_i A_j \partial_k f - \partial_i A_a A_j \partial_k f \Big)+ O(\Theta^3)\,.
}

\subsubsection*{Comments on an action}

After we have successfully derived the equations of motion for the
NC-CY theory up to order $\Theta^2$, one can ask whether these can be 
integrated to an action. Just to show what kind of  issues  appear, here we
consider
the simple case of the equations of motion up to linear order in
$\Theta$. Recall that the equation of motion is
\begin{eqnarray}
\label{hannover96}
           {\mathpzc F}&=&\ell_1(A) -{1\over 2} \ell_2(A,A) +O(\Theta^2)\\
     &=&\epsilon_c{}^{ab} \bigg( \partial_a A_b 
+{\textstyle {1\over 2}}\Theta^{ij} \partial_i A_a \partial_j A_b
    +\partial_a \Theta^{ij}  \Big(A_i \partial_j A_b 
    -{\textstyle {1\over 2}} A_i \partial_b A_j
    \Big)\bigg)+O(\Theta^2)\,. \nonumber
\end{eqnarray}
As mentioned in section \ref{sect_22}, for defining an action  one needs a
inner product, which we choose to be the same one as for the Moyal-Weyl case 
\eq{    
\langle A, E \rangle = \int d^3x \, \eta^{ab} A_a\, E_b\,.
}
For the action we also use the same form
\eq{
\label{actioncsone}
    S=  {\textstyle {1\over 2}} \langle A, \ell_1(A)
      \rangle -{1\over 3!} \langle A, \ell_2(A,A) \rangle +O(\Theta^2)\,.
}
Varying now this action with respect to the gauge field one has to do
a partial integration. In doing this, we require
\eq{
                 \partial_i \Theta^{ij} =0
}
which can be considered as the open string equation of motion for a
flat brane or  a natural  topological generalization of the latter.
Thus, after variation we can express the result as
\eq{
   \delta S=\int d^3 x\; \biggl[ &\delta A_c \;\epsilon^{abc} \Big(
                          \partial_a A_b  +{\textstyle {1\over
                              2}}\Theta^{ij} \partial_i A_a \partial_j
                          A_b \Big)+\\
                    & \delta A_c \;\epsilon^{abc} \partial_a
\Theta^{ij}\Big( A_i \partial_j A_b
-{\textstyle {1\over 2}} A_i \partial_b A_j\Big)-\\
&\delta A_c \;\epsilon^{abc} \partial_a
\Theta^{ij}\Big( {\textstyle {1\over 3}}  A_i (\partial_j A_b- \partial_b A_j)
-{\textstyle {1\over 6}} A_b (\partial_j A_i-\partial_i A_j) \Big)+\\
& \delta A_i \;\epsilon^{abc} \partial_a
\Theta^{ij}\Big( {\textstyle {1\over 3}} A_c (\partial_j
A_b-\partial_b A_j)-{\textstyle {1\over 12}} A_j (\partial_b
A_c-\partial_c A_b)\Big)\biggr]\,.
}
Let us make two remarks: First, the first four terms are  already the ones
appearing in the equation of motion. 
Second, from the variation one  gets  extra terms, e.g. those
multiplying $\delta A_i$. However, all these terms explicitly contain a
factor of  $\partial \Theta$ and are proportional 
to $\partial_k A_l-\partial_l A_k$. Since the leading
order equation of motion tells us that this combination vanishes, 
all the terms like $\partial  \Theta (\partial_k A_l-\partial_l A_k)$
are already of  order $\Theta^2$ and can be safely dropped from 
the equation of motion at first order in $\Theta$.
Therefore, only after carefully working at linear order in $\Theta$
one gets the equation of motion \eqref{hannover96} from the action
\eqref{actioncsone}. This means that the inner product  itself
does not satisfy the cyclicity property \eqref{cyclicprop}.

It would be interesting to see whether such a reasoning also works 
up to second order in $\Theta$ or whether also the inner product  receives
some derivative $\partial \Theta$ corrections. We leave this 
involved study for future research.

\subsection{NC Yang-Mills theory}

So far we have considered NC deformations of the topological
CS-theory. To really get into contact  with the D-branes appearing
as solutions of string theory,
we finally consider non-commutative $U(1)$ YM theory.
Since in this case,
the computations turn out to be more involved than in the CS case, we
restrict ourselves in this section only to linear order in $\Theta$
and leave the generalization to higher orders for the future.

Since the equation of motion involves the metric, we first discuss the
simplest case of  a flat background
with a non-constant NC-structure  $\Theta$ and then generalize this 
to a curved background of metric $g_{ij}$ and non-constant $\Theta$.

\subsubsection*{NC-YM on flat background}

Recall from section \ref{sec_CSYML} that for the Moyal-Weyl
star-product we have three vector spaces like
\eq{      
\begin{matrix}   X_0\quad  &\quad X_{-1}\quad  &\quad  X_{-2} \\[0.2cm]
                        f    \quad  &\quad   A_a   \quad  &\quad  E_a
\end{matrix}
}
with the following $\ell$-products
\begin{eqnarray}
     \ell_1(A)&=&\Box A_{a}-\partial_a(\pd\cdot A) \\
    \ell^{(0)}_2(A,B)&=&i [\partial\cdot A,B_a]_\star + i   [A_k,\partial^k B_a]_\star +i [A^k,\partial_k
    B_a-\partial_a B_k]_\star +(A \leftrightarrow B) \nonumber\\
    \ell_3(A,B,C)&=&[A^k,[B_k,C_a]_\star]_\star + {\rm 5\; terms}\,.\nonumber
\end{eqnarray}
Note that at linear order in $\Theta$ the three-product
$\ell_3(A,B,C)$ is vanishing. As for the NC-CS theory, imposing
the Leibniz-rule we expect to get a derivative correction
$\ell^{(1)}_2(A,B)$.
Indeed, going through the computation we arrive at the familiar expression
\eq{
             \ell_2(f,E)=i [f,E_a]_\star
}
and the remaining term
\eq{
    \ell^{(1)}_2(\ell_1(f),A)&=-\Box\Theta^{ij}\Big( \partial_i
    f \partial_j A_a - {\textstyle {1\over 2}} \partial_a\partial_i f
    A_j -{\textstyle {1\over 2}} \partial_i f
    \partial_a A_j\Big)\\[0.1cm]
    &\hspace{0.4cm} +\big(\partial^k \partial_a \Theta^{ij}\big)\Big(
       \partial_i f \partial_j A_k 
- {\textstyle {1\over 2}} \partial_i\partial_k f
    A_j -{\textstyle {1\over 2}} \partial_i f
    \partial_k A_j\Big)\\[0.1cm]
  &\hspace{0.4cm} -\big(\partial_a \Theta^{ij}\big)\Big( \partial^k\partial_i f
  (\partial_j A_k-\partial_k A_j) -\partial_i f \partial_j(\partial \cdot A)\\
 &\hspace{0.4cm} \phantom{aaaaaaaaaa}  +{\textstyle {1\over 2}} \partial_i f  \Box A_j 
+{\textstyle {1\over 2}} \Box\partial_i f   A_j\Big) \\[0.1cm]
&\hspace{0.4cm} -\big( \partial^k\Theta^{ij}\big)\Big( 2 \partial_k \partial_i
f \partial_j A_a +2  \partial_i f \partial_k\partial_j A_a \\
&\hspace{0.4cm} \phantom{aaaaaaaaaa}  -\partial_a \partial_i f \partial_j A_k-\partial_i f \partial_ a\partial_j A_k\\
&\hspace{0.4cm} \phantom{aaaaaaaaaa}  - {\textstyle {1\over 2}} \partial_a \partial_k \partial_i f A_j
- {\textstyle {1\over 2}} \partial_k\partial_i f \partial_a A_j\\
&\hspace{0.4cm} \phantom{aaaaaaaaaa}  - {\textstyle {1\over 2}} \partial_a\partial_i f \partial_k A_j
- {\textstyle {1\over 2}} \partial_i f \partial_a \partial_k A_j\Big)\,.
} 
Again, the relative coefficients are just right to be able to read off
a symmetric $\ell_2(A,B)$
\eqb{
    \ell_2(A,B)&=i [\partial\cdot A,B_a]_\star + i   [A_k,\partial^k B_a]_\star
   +i [A^k,\partial_k B_a-\partial_a B_k]_\star\\[0.1cm]
&\hspace{0.4cm} -\Box\Theta^{ij}\Big( A_i \partial_j B_a 
  - {\textstyle {1\over 2}} A_i \partial_a B_j \Big)\\[0.1cm]
&\hspace{0.4cm} +\big(\partial^k \partial_a \Theta^{ij}\big)\Big(
       A_i  \partial_j B_k 
- {\textstyle {1\over 2}} A_i \partial_k B_j  \Big)\\[0.1cm]
&\hspace{0.4cm} -\big(\partial_a \Theta^{ij}\big)\Big( \partial^k A_i 
  \partial_j B_k  -A_i  \partial_j(\partial \cdot B)
   +{\textstyle {1\over 2}} A_i  \Box B_j  \Big) \\[0.1cm]
&\hspace{0.4cm} -\big( \partial^k\Theta^{ij}\big)\Big( 
{\textstyle {3\over 2}}   \partial_i A_k \partial_j B_a 
 + {\textstyle {1\over 2}}  \partial_k A_i  \partial_j B_a 
+ {\textstyle {1\over 2}}  \partial_a A_i  \partial_j B_k \\
&\hspace{0.4cm}\phantom{aaaaaaaaaa}  + 2  A_i  \partial_k \partial_j B_a 
-    A_i  \partial_a \partial_j B_k
 - {\textstyle {1\over 2}}   A_i  \partial_a \partial_k  B_j \Big) \\[0.1cm]
&\hspace{0.4cm} \phantom{aaaaaaaaaaaaa} +(A\leftrightarrow B)\,.
 }
Then, the vacuum equation of motion for the non-commutative $U(1)$
Yang-Mills theory  up to linear order in $\Theta$ reads
\eq{
\label{YM3b}
0&=\Box A_{a}-\partial_a(\pd\cdot A)
-i [\partial\cdot A,A_a]_\star - i   [A_k,\partial^k A_a]_\star
-i[A^k,\partial_k
A_a-\partial_a A_k]_\star\\[0.1cm]
&\hspace{0.4cm} +\Box\Theta^{ij}\Big( A_i \partial_j A_a 
  - {\textstyle {1\over 2}} A_i \partial_a A_j \Big)\\[0.1cm]
&\hspace{0.4cm} -\big(\partial^k \partial_a \Theta^{ij}\big)\Big(
       A_i  \partial_j A_k 
- {\textstyle {1\over 2}} A_i \partial_k A_j  \Big)\\[0.1cm]
&\hspace{0.4cm} +\big(\partial_a \Theta^{ij}\big)\Big( \partial^k A_i 
  \partial_j A_k  -A_i  \partial_j(\partial \cdot A)
   +{\textstyle {1\over 2}} A_i  \Box A_j  \Big) \\[0.1cm]
&\hspace{0.4cm} +\big( \partial^k\Theta^{ij}\big)\Big( 
{\textstyle {3\over 2}}   \partial_i A_k \partial_j A_a 
 + {\textstyle {1\over 2}}  \partial_k A_i  \partial_j A_a 
+ {\textstyle {1\over 2}}  \partial_a A_i  \partial_j A_k \\
&\hspace{0.4cm} \phantom{aaaaaaaaaa}  + 2  A_i  \partial_k \partial_j A_a 
-    A_i  \partial_a \partial_j A_k
 - {\textstyle {1\over 2}}   A_i  \partial_a \partial_k  A_j \Big) \,.
}
As we see, the linear order corrections are much more involved than
for NC-CS theory so that we stop here and leave higher order
computations for future work. Here it is important to note that
so far we did not encounter any obstacle for solving the L$_\infty$
relations.

\subsubsection*{NC-YM on curved background}

As we have seen in section \ref{sec_Dbranes}, the general situation for on-shell
$D$-brane configurations involves a curved manifold equipped with
the (open string) metric $G$  and a non-constant bi-vector
$\Theta$.
On such a Riemannian  manifold the easiest case should be one where
$\Theta$ is covariantly constant with respect to the Levi-Civita
connection. This requirement is  natural from the open string equation of
motion \eqref{eomcovar} for the WZW branes. 
For instance, the two-dimensional
branes for the $SU(2)$ WZW model really feature a 
covariantly constant $\Theta$.

Clearly, here we are entering new territory, as
the usual star-product is constructed with respect to a Poisson
structure only, without any mentioning of a metric or a connection.
In this section, at least up to  linear order in $\Theta$,
 we investigate whether one can also bootstrap the first terms
of a NC-YM theory on a curved manifold following our strategy of imposing the
L$_\infty$ algebra.

Looking at the usual abelian Yang-Mills theory on a Riemannian manifold,
our  starting point is that 
\eq{
             \ell_1(f)=\partial_a f\,,\qquad 
             \ell_1(A)=\nabla^b \nabla_b A_a - \nabla^b \nabla_a A_b
}
where the second definition follows from varying the action
\eq{
             S=\int d^n x\, \sqrt{G}\, F_{ab}\, F^{ab}  
}
where indices are pulled up and down with the metric $G_{ab}$.
Moreover, for the star-product between two functions up to linear order we keep
\eq{
          f\star g = f\cdot g + {i\over 2} \Theta^{ij} \partial_i
          f \partial_j g + O(\Theta^2)
} 
and define
\eq{
            \ell_2(f,g)=i [f,g]_\star \,.
}
The Leibniz rule ${\cal J}_2(f,g)=0$ can now be used to bootstrap 
the form of $\ell_2(f,A)$, assuming that $\nabla_k \Theta^{ij}=0$
\eq{
      \ell_2(f,A)=i [f,A]_\star := -\Theta^{ij}\, \nabla_i f\, \nabla_j A_a + O(\Theta^2)\,
}       
where of course $\nabla_i f=\partial_i f$. Thus, we realize that from
this
perspective it is more natural that indeed the covariant derivative
appears when star-multiplying tensors. Since two covariant derivatives
do not commute and give
\eq{
           [\nabla_a,\nabla_b] T_c=R_{ab,c}{}^d\, T_d\,,
}
we must be prepared that there will arise 
curvature\footnote{Note that the torsion vanishes for the Levi-Civita connection.}
corrections to the expression we encountered in the previous sections.
Next, we impose the Leibniz rule  ${\cal J}_2(f,A)=0$ from which we
are  able to read-off $\ell_2(A,B)$ and $\ell_2(f,E)$.
After reordering covariant derivatives and applying  Bianchi-identities
for the curvature, we finally arrive at 
\eq{
               \ell_2(f,E)=i [f,E]_\star
}
and the more involved expression
\eq{
       \ell_2(A,B)&=\nabla^b \Big( i [A_b, B_a]_\star +{1\over 2}
       \Theta^{ij} R_{ab,j}{}^c \,A_i\, B_c + (A\leftrightarrow B
       )\Big)\\[0.1cm]
           &\phantom{=} +\Big( i [A^b, \partial_b B_a - \partial_a B_b]_\star 
               -\Theta^{ij} R_{jc}\, A_i\,  (\partial^c B_a - \partial_a
               B^c)\\
             &\phantom{=aaaa} - \Theta^{ij} R_{jb,a}{}^c \, A_i\,  (\partial_c B^b - \partial^b
               B_c) + (A\leftrightarrow B) \Big)\,.
}
Here, we have used $\partial_{[k} B_{l]}= \nabla_{[k} B_{l]}$.
Note that this expression has the correct limit in the flat case and
manifestly shows the curvature corrections.
Since all $\ell_2$-product are at first order in $\Theta$, all higher
order relations are satisfied up to linear order and we have succeeded
to bootstrap a NC-YM theory on a curved manifold up to linear order
in a covariantly constant $\Theta$.

This could be continued  to higher orders but we do not pursue this
further here and just state that all the discussed examples exemplify
that  the string theory motivated L$_\infty$ bootstrap program
provides a promising novel approach 
to algebraically construct non-commutative gauge theories in regimes
that were not completely accessible yet.

\section{Conclusions}

Motivated by its appearance in string theory and first successes when
applied to the matrix valued NC gauge theory on the fuzzy 2-sphere, in the main part
of  this paper we have successfully carried out an L$_\infty$  bootstrap program
for determining higher derivative corrections to  NC gauge theories
for non-constant and in general non-associative NC-structure $\Theta$.
What is changed is both, the action of the gauge symmetry on the
gauge fields and their equations of motion. An interesting open
question is whether one can also find an action for them. We leave
this non-trivial question for future work.

We believe  that this approach is different from former attempts to solve
this problem but is based on a string theoretic well motivated guiding
principle. By successively applying or solving the L$_\infty$
relations we managed to determine higher  $\ell$-products.
Since we were pursuing a perturbative approach in $\Theta$, the actual
computations  become more and more involved. 
Note that at higher orders also the Kontsevich
star-commutator  receives derivative corrections that one needs
to take into account. (For the A$_\infty$ algebra from appendix
\ref{app_B} such corrections already appeared at second order and
could consistently be handled.)
But it is promising that, up to second order in $\Theta$, we did not encounter any obstacle,
even not in the non-associative case. To gain even more evidence for the
self-consistency  of this approach, one could try to implement this
bootstrap algorithm and push the computer to iteratively produce
higher and higher orders. Mathematically, one could also ask
for a proof that our algorithm always works.

When considering NC-YM theory on a curved space with covariantly
constant
$\Theta$, the L$_\infty$ structure was telling us that one should
better introduce a star-product that uses covariant derivatives when
acting on tensors. It would be interesting to investigate whether
our first order results can be extended to higher orders.

Of course, generally one could contemplate on other possible problems
where such an L$_\infty$  based bootstrap approach might be worthwhile to pursue.

\vspace{0.8cm}

\noindent
\subsubsection*{Acknowledgments}
We would like to thank Max Brinkmann, Andreas Deser, Michael Fuchs, Matthias Traube  and Barton
Zwiebach for discussion. 
We also acknowledge 
that this project was strongly influenced by
the nice atmosphere at the
{\it Workshop on  Noncommutative Field Theory and Gravity} at the
Corfu Summer Institute 2017. 
The work of D.L.~  is supported by the
ERC Advanced Grant No.~320045 ``Strings and Gravity". I.B. and D.L. are supported by the Excellence Cluster Universe. V.K. acknowledges the CAPES-Humboldt Fellowship No.~0079/16-2 and CNPq Grant No.~305372/2016-5.

\clearpage
\appendix


\section{Semi classical branes in  $SU(3)$ WZW} 
\label{app_A}

As we have seen, the $SU(2)$ group manifold is not rich enough to
non-trivially check whether on-shell brane configurations exist with
non-vanishing $\Pi^{ijk}$. The reason was that the boundary states 
describe at most  two-dimensional branes on which $H$ restricts trivially.
Thus, we now consider the eight-dimensional $SU(3)$ 
WZW model which admit six-dimensional branes. The non-vanishing Betti numbers of this group
manifold are 
\eq{
            b^0=b^3=b^5=b^8=1\, .
}
This manifold can be considered as a $S^3$ fibration over a
five-dimensional base $M_5$. Clearly, in order to satisfy the Freed-Witten
anomaly condition, the $D$-brane world-volume should better not
contain the $S^3$.

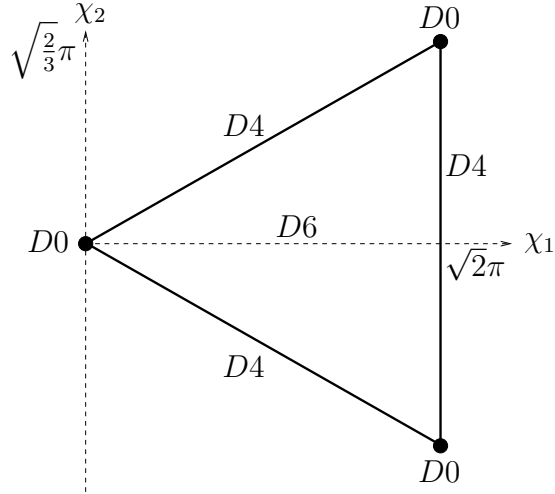
\begin{figure}[ht]
  \centering
  \includegraphics[width=0.4\textwidth]{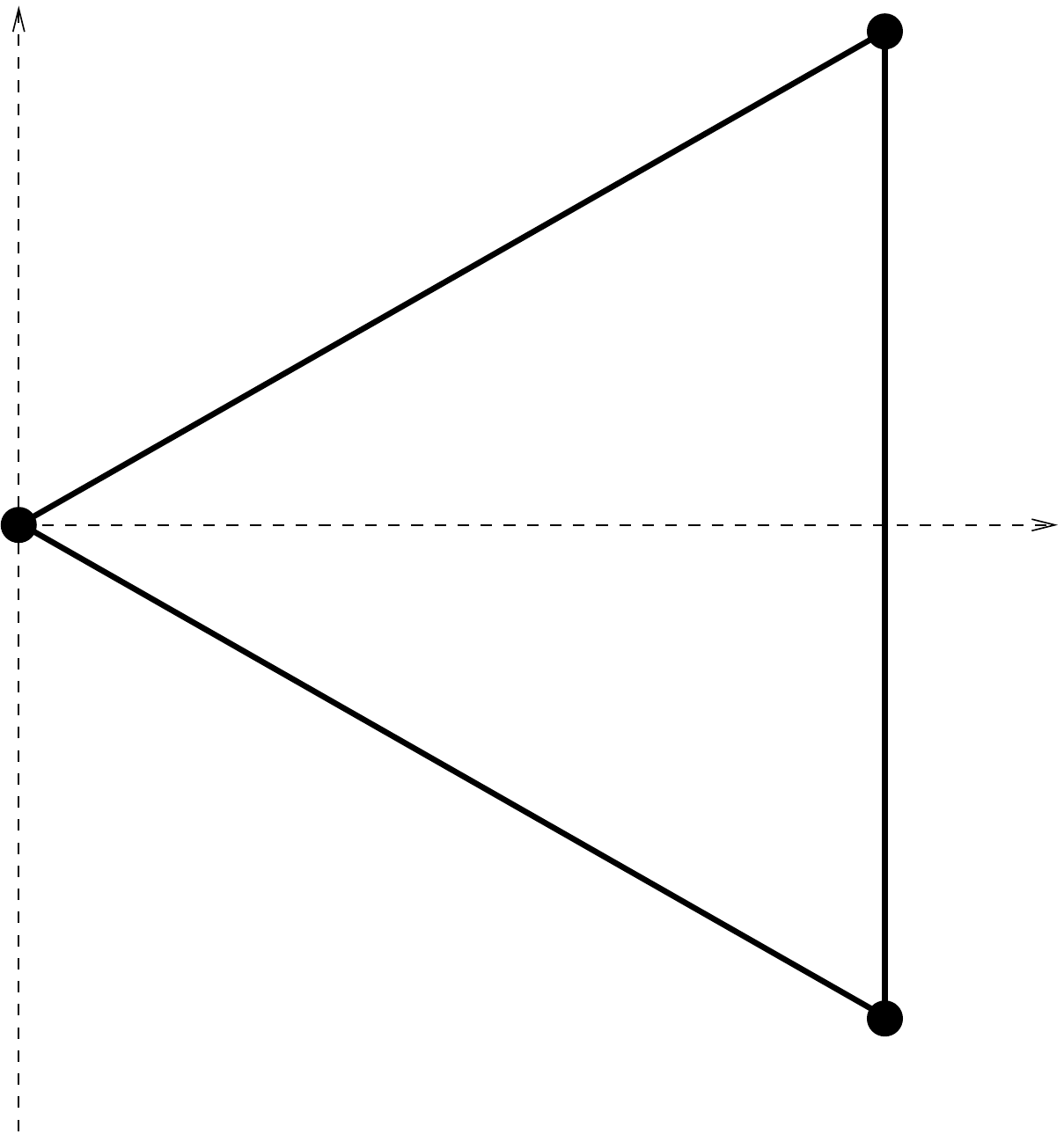}
\begin{picture}(0,0)
    \put(-170,180){$\chi_2$}
   \put(0,93){$\chi_1$}
   \put(-30,82){$\sqrt 2 \pi$}
   \put(-195,164){$\textstyle{\sqrt{2\over 3}} \pi$}
    \put(-94,97){$D6$}
    \put(-114,135){$D4$}
    \put(-114,44){$D4$}
    \put(-30,120){$D4$}
     \put(-188,91){$D0$}
    \put(-40,4){$D0$}
    \put(-40,176){$D0$}
    \end{picture}
  \caption{Domain for position of the $D6$-brane.}
  \label{fig:domain}
\end{figure}

From these topological considerations
this example seems to be rich enough to provide a non-trivial example
with $\Pi^{ijk}\ne 0$.
To explicitly describe the $SU(3)$ group manifold we introduce 
the matrices
\eq{
D_1(\chi_1)=\left(\begin{matrix}  e^{{i\over \sqrt{2}} \chi_1} & 0 & 0 \\
                   0 & e^{-{i\over \sqrt{2}} \chi_1}  & 0  \\
                        0 & 0 & 1 \end{matrix}\right), \ 
D_2(\chi_2)=\left(\begin{matrix}   e^{{i\over \sqrt{6}} \chi_2} & 0 & 0 \\
                   0 &  e^{{i\over \sqrt{6}} \chi_2}  & 0  \\
                        0 & 0 &    e^{-i \sqrt{2\over 3} \chi_2} \end{matrix}\right)
}
and
\eq{
M_{12}(\varphi,\psi)=\left(\begin{matrix}  \cos \varphi & \sin \varphi
    \, e^{i\psi} & 0 \\
          -\sin \varphi \, e^{-i\psi}  & \cos \varphi &0 \\
              0 & 0 & 1 \end{matrix}\right)
}
and similarly for $M_{13}$ and $M_{23}$.
Now we write an element of $SU(3)$ as
\eq{
         M= N^{-1}\, D_1(\chi_1)\, D_2(\chi_2)\, N
}
with $N=M_{12}(\varphi_1,\psi_1)\, M_{23}(\varphi_2,\psi_2)\, M_{12}(\varphi_3,\psi_3)$. 
Thus, the six coordinates $\phi_i$ along the brane are $\phi\in\{\varphi_1,\psi_1,\varphi_2,\psi_2,\varphi_3,\psi_3\}$
Moreover, in this normalization the positive roots are given by
\eq{
            \alpha_1=(\sqrt 2,0)\,,\quad \alpha_2=\Big({\textstyle {1\over \sqrt
              2},\sqrt{3\over 2}}\Big) \,,\quad
          \alpha_3=\Big({\textstyle -{1\over \sqrt
              2},\sqrt{3\over 2}}\Big)\,.
}
For the metric on $SU(3)$ one obtain
\eq{
             k^{-1}\,   ds^2={1\over 2} d\chi_1^2 +{1\over 2} d\chi_2^2 +
                g_{ij}(\phi)\, d\phi^i\otimes d\phi^j
}
where  the second term is the metric restricted to the $D$-brane.
The metric components $g_{ij}(\phi)$ are partially long expressions
in terms of the coordinates along the brane.
Therefore, we just list a few components to  convince the reader
that the expressions are indeed very explicit
\eq{
           g_{11}&=4 \sin^2\left({\chi_1\over \sqrt 2}\right)\,,\qquad g_{12}=g_{13}=g_{14}=0\\
           g_{15}&=4 \sin^2\left({\chi_1\over \sqrt 2}\right)
           \cos\varphi_2 \cos(\psi_1-\psi_3) \\
           g_{16}&=2 \sin^2\left({\chi_1\over \sqrt 2}\right)
           \cos\varphi_2 \sin(2\varphi_3) \sin(\psi_1-\psi_3) \\
            g_{22}&= \sin^2\left({\chi_1\over \sqrt 2}\right) \sin^2(2\varphi_1)\,, \ldots
} 
At generic positions $\chi_{1,2}$ this gives a smooth metric on the
$D6$-brane. However, at the three boundaries $\alpha_i\cdot \chi =0$ 
the metric degenerates to a four-dimensional metric. Therefore, at
these positions one gets $D4$-branes. At the intersection of two such
lines the whole metric degenerates thus yielding
$D0$-branes. Therefore, the position moduli space has the form 
displayed in figure \ref{fig:domain}.
\vspace{0.3cm}

\noindent
The determinant of the metric  has a simple form
\eq{
           \sqrt{g}=4k^4\,\prod_{\alpha>0} \sin^2\!\left( \textstyle{
               {\alpha\cdot\chi \over 2}}\right) \,\sin(2\varphi_1)
         \sin(2\varphi_2) \sin^2(\varphi_2) \sin(2\varphi_3)\,.
}
Integrating this over the domain
\eq{
         0\le \chi_1\le \sqrt{2}\pi\,,\quad   -{\chi_1\over \sqrt{3}}\le \chi_2\le
         {\chi_1\over \sqrt{3}} \,,\quad
           0\le\varphi_i\le {\pi\over 2}\,,\quad   0\le\psi_i\le 2\pi
}
one finds for the volume of the $SU(3)$ group manifold
\eq{
                V(SU(3))=\sqrt{3} \pi^5 k^4\,.
}

Next, utilizing \eqref{bfield_brane} one computes the $B$-field  and its total derivative to get the
$H$-flux.  To find a compact expression for the sechs-bein along the D-brane
world-volume it is useful to define the structure ''constants''
\eq{
                      F_{AB}{}^C(\phi):=2\,\hat\theta_{[A}{}^i \,\partial_i
                      \hat\theta_{B]}{}^j\, \theta^C{}_j
}
where $A,B,C\in\{ \alpha_i,\ov\alpha_i\}$.
There are  non-zero and non-constant elements $F_{AB}{}^A(\phi)$ (no sum over $A$), but they do not
contribute to $H$.  The really relevant constant non-zero elements turn out to be
\eq{
   F_{31}{}^2=F_{\ov 3\ov 1}{}^{\ov 2}=F_{\ov 1 2}{}^3=F_{\ov 2
     3}{}^{\ov 1}=F_{1\ov 2}{}^{\ov 3}=F_{2\ov 3}{}^1=1\,.
} 
Now, using
\eq{
                     d\theta^A=-{1\over 2} F_{BC}{}^A \,\theta^B\wedge \theta^C
}
 one can  express the $H$-flux  in a very compelling  way as
\eq{
    H\vert_{\rm D}=h(\chi)
\,\Big[\theta^{\alpha_1}\wedge
    \theta^{\ov\alpha_2}\wedge  \theta^{\alpha_3} - \theta^{\ov\alpha_1}\wedge
    \theta^{\alpha_2}\wedge  \theta^{\ov\alpha_3}\Big]
 }
with
\eq{
                h(\chi)&=i k \sum_{i=1}^3  (-1)^i\Big( \alpha_i\cdot \chi -\sin(\alpha_i\cdot \chi)\Big)\\
           &=   2ik    \sin\left( \textstyle{ {\chi_1\over \sqrt 2}}\right)\!
    \left[ \cos\left( \textstyle{{\chi_1\over \sqrt
       2}}\right)-\cos\left(\textstyle{\sqrt{3\over 2}
     {\chi_2}}\right) \right]\,.
}
Thus, in contrast to the former $SU(2)$ case, the restriction of
$H$ onto the $D6$-brane is not vanishing. However, the restriction
onto the $D4$ and $D0$ branes at boundary of the moduli space
vanishes.
Due to  $\sum_i  (-1)^i \alpha_i\cdot \chi=0$ these linear terms in $\chi$ do
not contribute to  the three-form flux $H\vert_{\rm D}$ on the brane.
Taking into account \eqref{gaugeflux}, this is nothing else than the
manifestation 
of the fact  that a pure gauge flux $F=dA$
satisfies $dF=0$.  Therefore, in accord with the Freed-Witten anomaly,
the restriction of the bulk $H$-flux
onto the brane can be expressed as $H\vert_{\rm D}=d{\cal F}$. 
Being non-vanishing,  there is a good chance to finally also get a non-vanishing  $\Pi^{ijk}$. 
 
The quantization of the gauge flux $F$ only admits a finite number of
allowed $D$-branes. These are parametrized by 
\eq{
                 \chi_1= \sqrt{2}\pi\, {m_1\over k}\,,\quad
                 \chi_2={\textstyle \sqrt{2\over 3}}\pi \, {(2m_2-m_1)\over k} 
}
with $m_1,m_2\in\mathbb Z$ such that they lie inside the domain 
in figure \ref{fig:domain}. For $k=1$ one only gets the three
$D0$-branes but for $k=3$, for the first time, also a $D6$ is allowed.

Next one can derive an explicit expression of the flux ${\cal F}$ and 
compute 
\eq{ \sqrt{g+{\cal F}}= {k^3\over 2} \prod_{\alpha>0} \sin\!\left( \textstyle{
               {\alpha\cdot\chi \over 2}}\right) \,\sin(2\varphi_1)
         \sin(2\varphi_2) \sin^2(\varphi_2) \sin(2\varphi_3)\,.
}
Similarly, from \eqref{bivector} one can get an  expression for the anti-symmetric bi-vector
$\Theta$ and explicitly check 
that the equation of motion \eqref{openeom} is indeed
satisfied. Starting with the general expression  \eqref{bivector}, one
can check in more detail that the underlying reason for this result  are the relations
\eq{
                   \partial_i \left( \sqrt{g+{\cal F}} \,
                     \hat\theta_\alpha{}^i\right)&=- \sqrt{g+{\cal F}}
                   \, F_{\alpha\ov\alpha}{}^{\ov\alpha}(\phi)\\
      \partial_i \left( \sqrt{g+{\cal F}} \,
                     \hat\theta_{\ov\alpha}{}^i\right)&=- \sqrt{g+{\cal F}}
                   \, F_{\ov\alpha\alpha}{}^{\alpha}(\phi)
}
(no sum over $\alpha$ in $F_{\ov\alpha\alpha}{}^{\alpha}$)
leading to a cancellation already for each term in the sum over the positive
roots in \eqref{bivector}.
Thus, as expected,
the highly curved, fluxed $D6$-brane is  a consistent solution of the string equations of motion.
Similarly to the $H$-flux one can write the non-vanishing antisymmetric
three-vector  $\Pi=[\Theta,\Theta ]_S$ in the
very compelling form 
\eq{
\setlength\fboxsep{0.3cm}
\setlength\fboxrule{0.5pt}
\boxed{
    \Pi={k^2\over 4} \pi(\chi)    
    \,\Big[\hat\theta_{\alpha_1}\wedge
    \hat\theta_{\ov\alpha_2}\wedge  \hat\theta_{\alpha_3} +\hat\theta_{\ov\alpha_1}\wedge
    \hat\theta_{\alpha_2}\wedge  \hat\theta_{\ov\alpha_3}\Big]
 }
}
with 
\eq{
         \pi(\chi)&=\cot\left(\textstyle{{\alpha_1\cdot\chi\over
             2}}\right)\cot\left(\textstyle{{\alpha_2\cdot\chi\over 2}}\right)-
       \cot\left(\textstyle{{\alpha_1\cdot\chi\over
             2}}\right)\cot\left(\textstyle{{\alpha_3\cdot\chi\over
             2}}\right)+
       \cot\left(\textstyle{{\alpha_2\cdot\chi\over
             2}}\right)\cot\left(\textstyle{{\alpha_3\cdot\chi\over
             2}}\right)\\[0.2cm]
     &=-1\,.
}
This explicitly shows that the $SU(3)$ WZW
model admits a six-dimensional brane that carries a non-trivial $\Pi$.
That means that $\Theta$ is not a Poisson structure and the
related star-product becomes non-associative.
For concreteness, we display  the components of $\Pi$
\eq{
\Pi^{123}&={\sin(\psi_1-\psi_3)\, \sin\phi_3\over 8 \sin\phi_2
  \cos\phi_3}\\
\Pi^{124}&={\cos(2\phi_1) \over 4 \sin(2\phi_1) \sin^2(\phi_2)}-
{\cos(\psi_1-\psi_3) (5+3\cos(2\phi_2)) \sin\phi_3\over 32 \sin^2 (\phi_2)
  \cos\phi_2 \cos \phi_3}\\
\Pi^{134}&=- {\sin(\psi_1-\psi_3)\, \sin\phi_3\over 16 \sin\phi_2
  \cos\phi_3}\\
\Pi^{234}&=-{1\over 4 \sin(2\phi_2) }-
{\cos(\psi_1-\psi_3) \cos(2\phi_1) \sin\phi_3\over 8 \sin(2\phi_1)
  \sin\phi_2 \cos \phi_3}\\
\Pi^{125}&={\sin(\psi_1-\psi_3) (1+3\cos(2\phi_2))\over 32
  \sin^2(\phi_2)  \cos\phi_2 }\\
\Pi^{135}&=- {\cos(\psi_1-\psi_3) \over 16 \sin\phi_2}\\
\Pi^{235}&={\sin(\psi_1-\psi_3) \cos(2\phi_1)  \over 16 \sin\phi_1
  \cos\phi_1 \sin\phi_2}\\
\Pi^{145}&=-{\sin(\psi_1-\psi_3) \over 8 \sin\phi_2
  \sin(2\phi_2)}\\
\Pi^{245}&={\sin(\phi_3) \over 8 \sin^2(\phi_2)\cos\phi_3}-
{\cos(\psi_1-\psi_3) \cos(2\phi_1)\over 8 \sin(2\phi_1) \sin^2(\phi_2) \cos \phi_2}\\
\Pi^{345}&=0\\
\Pi^{126}&=-{\cos(\psi_1-\psi_3) (1+3\cos(2\phi_2))\over 16
  \sin^2(\phi_2)  \cos\phi_2 \sin(2\phi_3)}\\
\Pi^{136}&=-{\sin(\psi_1-\psi_3) \over 8 \sin\phi_2
  \sin(2\phi_3)}\\
\Pi^{236}&=-{\cos(\psi_1-\psi_3) \cos(2\phi_1)\over 4
  \sin(2\phi_1)  \sin\phi_2\sin(2\phi_3)}\\
\Pi^{146}&={\cos(\psi_1-\psi_3)  \over 16 \cos\phi_2 
  \sin^2(\phi_2) \sin\phi_3 \cos\phi_3}\\
\Pi^{246}&=-{\sin(\psi_1-\psi_3)  \cos(2\phi_1)\over 2 \sin(2\phi_1) \sin\phi_2 
  \sin(2\phi_2) \sin(2\phi_3) }\\
\Pi^{346}&=\Pi^{156}=0\\
\Pi^{256}&=-{1\over 8 \sin^2(\phi_2) \sin\phi_3 \cos\phi_3}\\
\Pi^{356}&=\Pi^{456}=0\\
}

\section{Refinement to an A$_\infty^{\rm gauge}$ algebra}
\label{app_B}

Since in string theory, open-closed string field theory has an
underlying A$_\infty$ algebra, one might expect that 
 NC-gauge theory admits an  A$_\infty$ algebra, as well.
Thus, in this appendix we refine the L$_\infty^{\rm gauge}$ structure
from section 4 and construct an  A$_\infty^{\rm gauge}$ algebra that underlies  the,
in general non-associative, gauge theory up to order  $O(\Theta^2)$.

Recall that we consider two non-trivial vector spaces, $X_0$ and $X_{-1}$ where
$X_0$ contains functions (gauge parameters) and $X_{-1}$ vectors
(gauge fields). 
Moreover, to simplify the notation, in this section
we use 
\eq{
\hat\Theta^{ij}={i\over 2} \Theta^{ij}\,,\qquad 
\hat\Pi^{ijk}=-{1\over 4} \Pi^{ijk}.
}
Similar to the L$_\infty$ case, the first order products are defined
as\footnote{Like in \eqref{defm1m2}, in the following the
definitions of the higher products are put in boxes.}
\eqb{
\label{defm1m2}
   m_1(f)&=\partial_a f\in X_{-1}  \\
    m_2(f,g)&=f\star g +{1\over 3} (\hat\Theta^{im} \partial_m \hat\Theta^{j k})
    (\partial_i \partial_j f \,\partial_k g + \partial_i \partial_j
    g \,\partial_k f)\in X_0
}
where the second line is just the full Kontsevich star-product and 
$f\star g$ denotes the  Moyal-Weyl part of it
\eq{
     f\star g =f g +\hat\Theta^{ij} \partial_i f \partial_ g +{1\over 2}
     \hat\Theta^{im}\hat\Theta^{jn} \, \partial_i\partial_j
     f\, \partial_m\partial_n   g +O(\hat\Theta^3)\,.
}
By this split, we explicitly take care of all appearing derivative
$\partial \Theta$ terms.  
Note that $m_2(f,g)$ is neither symmetric nor anti-symmetric under
exchanging the arguments $f$ and $g$. The relation to the
corresponding higher
product in the L$_\infty$ algebra  is given by graded symmetrization
\eq{
             -i\ell_2(f,g)=m_2(f,g)-m_2(g,f)\,.
} 
The A$_\infty$ relation
\eq{
       {\cal A}_2(f,g)= m_1 (m_2(f,g) ) - m_2 (m_1(f),g)-m_2 (f,m_1(g))=0
}
is nothing else than the Leibniz-rule for the two-product and 
can be satisfied, up to order $O(\hat\Theta^2)$, by defining
\eqb{
m_2(A,g)&=A_a \star g +{1\over 3} (\hat\Theta^{im} \partial_m \hat\Theta^{j k})
    (\partial_i \partial_j A_a \,\partial_k g + \partial_i \partial_j
    g \,\partial_k A_a) \\
   &+{1\over 2} \partial_a\hat\Theta^{ij}\, A_i \star \partial_j
    g +{1\over 3} \partial_a( \hat\Theta^{im} \partial_m \hat\Theta^{jk}
    ) \,\partial_j A_i \,\partial_k g \\[0.1cm]
m_2(f,A)&= f\star A_a  +{1\over 3} (\hat\Theta^{im} \partial_m \hat\Theta^{j k})
    (\partial_i \partial_j f \,\partial_k A_a + \partial_i \partial_j
    A_a \,\partial_k f) \\
   &+{1\over 2} \partial_a\hat\Theta^{ij}\, \partial_i f \star A_j  
   +{1\over 3} \partial_a( \hat\Theta^{im} \partial_m \hat\Theta^{jk}
    ) \,\partial_j  A_i \,\partial_k f\,.
}
Note that the two terms in the second line are correction terms that
arise for non-constant  $\hat\Theta$.  These terms make the construction
of the A$_\infty$ algebra much more complicated than in the Moyal-Weyl
case of constant $\hat\Theta$. Again, by anti-symmetrization we can
confirm
\eq{   
                  -i\ell_2(f,A)=m_2(f,A)-m_2(A,f)\,.
} 
Next we analyze the A$_\infty$ relation
\eq{
\label{Arela}
       {\cal A}_3(f,g,h)&=  m_2(m_2(f,g),h)-m_2(f,m_2(g,h))+m_1(m_3(f,g,h))\\
    &+m_3(m_1(f),g,h)+m_3(f,m_1(g),h)+m_3(f,g,m_1(h))=0\,.
}
For the associator one finds
\eq{
\label{nonassopre}
           m_2(m_2(f,g),h)-m_2(f,m_2(g,h))=-{2\over 3}
           \hat\Pi^{ijk} \partial_i f \partial_j g \partial_k h\,.
}
Since $m_1(m_3(f,g,h))\in X_{1}$ vanishes one can solve \eqref{Arela}
by
\eqb{
   m_3(A,f,g)&={\alpha}\, \hat\Pi^{ijk} A_i \partial_j f \partial_k g\\
   m_3(f,A,g)&={\beta} \,\hat\Pi^{ijk} A_i \partial_j f \partial_k g \\
  m_3(f,g,A)&={\gamma}\, \hat\Pi^{ijk} A_i \partial_j f \partial_k g 
}
with $\alpha-\beta+\gamma=2/3$.
This guarantees that in the associative case, all these three-products
vanish.  It is straightforward to confirm the relation to the
corresponding L$_\infty$ three-product
\eq{
                  (-i)^2
                  \ell_3(A,f,g)&=m_3(A,f,g)-m_3(A,g,f)-m_3(f,A,g)+m_3(g,A,f)\\
                  &\phantom{=}+m_3(f,g,A)-m_3(g,f,A)\,.
}
Now we proceed invoking the remaining ${\cal A}_3$ relations
\eq{
\label{nonasso}
   {\cal A}_3(A,&f,g)=   m_2(m_2(A,f),g)-m_2(A,m_2(f,g))+m_1(m_3(A,f,g))\\
    &+m_3(m_1(A),f,g)-m_3(A,m_1(f),g)-m_3(A,f,m_1(g))=0\\[0.2cm]
   {\cal A}_3(f,&g,A)= m_2(m_2(f,g),A)-m_2(f,m_2(g,A))+m_1(m_3(f,g,A))\\
    &+m_3(m_1(f),g,A)+m_3(f,m_1(g),A)+m_3(f,g,m_1(A))=0\\[0.2cm]
   {\cal A}_3(f,&A,g)=m_2(m_2(g,A),f)-m_2(g,m_2(A,f))+m_1(m_3(g,A,f))\\
    &+m_3(m_1(g),A,f)+m_3(g,m_1(A),f)-m_3(g,A,m_1(f))=0\,.
}
Note the signs appearing in front of the $m_3\, m_1$-terms, which
involve extra signs relative to \eqref{relaerst} whenever $m_1$ is permuted
through an odd element $A\in X_{-1}$.
Let us first compute the associators
\eq{
      &m_2(m_2(A,f),g)-m_2(A,m_2(f,g))=\\
&\Big({1\over 3}  \hat\Theta^{im}\partial_a \partial_m \hat\Theta^{jk}
-{1\over 2}  \hat\Theta^{km}\partial_a \partial_m \hat\Theta^{ij}
-{1\over 6}  \partial_a \hat\Theta^{im} \partial_m \hat\Theta^{jk}
-{1\over 4}  \partial_a \hat\Theta^{km} \partial_m \hat\Theta^{ij}\Big)
A_a \partial_j f \partial_k g \\
&-{2\over 3} \hat\Pi^{ijk} \partial_i A_a \partial_j f \partial_k
g\\[0.2cm]
 &m_2(m_2(f,g),A)-m_2(f,m_2(g,A))=\\
&\Big({1\over 3}  \hat\Theta^{im}\partial_a \partial_m \hat\Theta^{jk}
-{1\over 2}  \hat\Theta^{jm}\partial_a \partial_m \hat\Theta^{ki}
-{1\over 6}  \partial_a \hat\Theta^{im} \partial_m \hat\Theta^{jk}
-{1\over 4}  \partial_a \hat\Theta^{jm} \partial_m \hat\Theta^{ki}\Big)
A_a \partial_j f \partial_k g \\
&-{2\over 3} \hat\Pi^{ijk} \partial_i A_a \partial_j f \partial_k
g}
and
\eq{
      m_2&(m_2(g,A),f)-m_2(g,m_2(A,f))=\\
\Big( &-{2\over 3}  \hat\Theta^{im}\partial_a \partial_m \hat\Theta^{jk}
-{1\over 2}  \hat\Theta^{km}\partial_a \partial_m \hat\Theta^{ij}
-{1\over 2}  \hat\Theta^{jm}\partial_a \partial_m \hat\Theta^{ki}\\
&-{2\over 3}  \partial_a \hat\Theta^{im} \partial_m \hat\Theta^{jk}
-{1\over 4}  \partial_a \hat\Theta^{km} \partial_m \hat\Theta^{ij}
-{1\over 4}  \partial_a \hat\Theta^{jm} \partial_m \hat\Theta^{ki}
\Big)
A_a \partial_j f \partial_k g \\
&-{2\over 3} \hat\Pi^{ijk} \partial_i A_a \partial_j f \partial_k
g\,.
}
To solve \eqref{nonasso}, we first observe that $m_1(A)=0$ and 
make the general ansatz
\eq{
         m_3(A,B,f)=\Sigma^{ijk} A_i B_j \partial_k f + \hat\Pi^{ijk} \Big(&
   x_1\, \partial_i A_a B_j \partial_k f + x_2\, A_i
   \partial_j B_a \partial_k f \\
   &+x_3\, \partial_a A_i B_j \partial_k f
+  x_4\, A_i  \partial_a B_j \partial_k f\Big)
}
and similarly for $m_3(A,f,B)$ and $m_3(f,A,B)$.
This gives a set of conditions that only admit a solution iff 
$\alpha-\beta+\gamma=2/3$, i.e. precisely the condition that 
followed from the relation ${\cal A}_3(f,g,h)=0$ shown in  \eqref{Arela}. Eliminating
$\beta$ in favor of $\alpha$ and $\gamma$, the general set of
solutions is given as
\eqb{
    m_3(A,&B,f)=\\
\bigg[ &\kappa_1 \, \hat\Theta^{im}\partial_a \partial_m \hat\Theta^{jk}
 +\kappa_2\,  \hat\Theta^{km}\partial_a \partial_m \hat\Theta^{ij}
 +\kappa_3\,  \hat\Theta^{jm}\partial_a \partial_m \hat\Theta^{ki}\\
 &+\lambda_1 \, \partial_a \hat\Theta^{im} \partial_m \hat\Theta^{jk}
 +\lambda_2\, \partial_a \hat\Theta^{km} \partial_m \hat\Theta^{ij}
 +\lambda_3 \,\partial_a \hat\Theta^{jm} \partial_m \hat\Theta^{ki}
\bigg] \,A_i B_j \partial_k f \\
+ \hat\Pi^{ijk} \bigg[&
   x_1\, \partial_i A_a B_j \partial_k f + x_2\, A_i
   \partial_j B_a \partial_k f +\\
   & (-{\textstyle {2\over 3}}+\alpha+\gamma-x_1)\, \partial_a A_i B_j \partial_k f
+  (\alpha-x_2)\, A_i  \partial_a B_j \partial_k f \bigg] \nonumber
}
and
\eqb{
    m_3(A,&f,B)=\\
\bigg[ & (-{\textstyle {1\over 3}} -\alpha+\kappa_1)\, \hat\Theta^{im}\partial_a \partial_m \hat\Theta^{jk}
 +(-\alpha+\kappa_3) \,  \hat\Theta^{km}\partial_a \partial_m \hat\Theta^{ij}\\
 &+ ({\textstyle {1\over 2}} -\alpha+\kappa_2)\,
 \hat\Theta^{jm}\partial_a \partial_m \hat\Theta^{ki} \bigg]\, A_i B_j \partial_k f \\
 +\bigg[ &({\textstyle {1\over 6}} -\alpha+\lambda_1 ) \, \partial_a \hat\Theta^{im} \partial_m \hat\Theta^{jk}
 + (-\alpha+\lambda_3) \,\partial_a \hat\Theta^{km} \partial_m \hat\Theta^{ij}\\
 &+({\textstyle {1\over 4}} -\alpha+\lambda_2)  \,\partial_a \hat\Theta^{jm} \partial_m \hat\Theta^{ki}
\bigg] \,A_i B_j \partial_k f \\
+ \hat\Pi^{ijk} \bigg[&
     ({\textstyle {2\over 3}}+x_1)\, \partial_i A_a B_j \partial_k f 
    + y_2\, A_i \partial_j B_a \partial_k f +\\
   & (-{\textstyle {2\over 3}}+\gamma-x_1)\, \partial_a A_i B_j \partial_k f
+  (-\alpha-y_2)\, A_i  \partial_a B_j \partial_k f\bigg]
}
and
\eqb{
    m_3(f,&A,B)=\\
\bigg[ &(-\alpha-\gamma+\kappa_3 )\, \hat\Theta^{im}\partial_a \partial_m \hat\Theta^{jk}
 +({\textstyle {1\over 6}} -\alpha-\gamma+\kappa_1)\,  \hat\Theta^{km}\partial_a \partial_m \hat\Theta^{ij}\\
 &+ ({\textstyle {1\over 6}} -\alpha-\gamma+\kappa_2)\,
 \hat\Theta^{jm}\partial_a \partial_m \hat\Theta^{ki} \bigg]\, A_i B_j \partial_k f \\
 +\bigg[ &(-\alpha-\gamma+\lambda_3 ) \, \partial_a \hat\Theta^{im} \partial_m \hat\Theta^{jk}
 +({\textstyle {5\over 12}} -\alpha-\gamma+\lambda_1) \,\partial_a \hat\Theta^{km} \partial_m \hat\Theta^{ij}\\
 &+({\textstyle {5\over 12}} -\alpha-\gamma+\lambda_2) \,\partial_a \hat\Theta^{jm} \partial_m \hat\Theta^{ki}
\bigg] \,A_i B_j \partial_k f \\
+ \hat\Pi^{ijk} \bigg[&
     (-{\textstyle {2\over 3}}+x_2)\, \partial_i A_a B_j \partial_k f 
    + ({\textstyle {2\over 3}}+y_2)\, A_i \partial_j B_a \partial_k f +\\
   & ({\textstyle {2\over 3}}-\gamma-x_2)\, \partial_a A_i B_j \partial_k f
+  (-\alpha-\gamma-y_2)\, A_i  \partial_a B_j \partial_k f\bigg] \nonumber
}
where besides $\alpha, \gamma$ the $\kappa_i$, $\lambda_i$ and
$x_1,x_2,y_2$ are still free parameters. However, when computing the
graded  symmetrization of these $m$-products, all these parameters
precisely cancel and one gets the corresponding $\ell_3$-product
\eq{
                 (-i)^2
                 \ell_3(A,B,f)&=m_3(A,B,f)+m_3(B,A,f)-m_3(A,f,B)\\
                      &\phantom{=} -m_3(B,f,A)+m_3(f,A,B)+m_3(B,A,f)\,.
}
Finally, one has to check the A$_\infty$ relation ${\cal A}_4$  \eqref{Arelationfour}.
There are only two possible
sets of a priori non-trivial relations with entries ${\cal A}_4(f,g,h,A)$ and
${\cal A}_4(f,g,A,B)$ and permutations thereof. The ${\cal A}_4(f,g,h,A)$ 
relations are all satisfied up to order  $O(\Theta^2)$ so that we
choose a vanishing four-product $m_4(f,g,A,B)\in X_0$.
The ${\cal A}_4(f,g,A,B)$ relations are also all satisfied in the
associative case, but in the non-associative case, one 
needs to introduce non-trivial  four-products $m_4(f,A,B,C)\in X_{-1}$ that are 
proportional to $\hat\Pi^{ijk}$. As before, we make a general ansatz
\eqb{
   m_4(f,A,B,C)=&\hat\Pi^{ijk} \Big( \mu_1 A_a B_i C_j +\mu_2 A_j B_a C_i
     +\mu_3 A_i B_j C_a \Big) \partial_k f \\
   +&\hat\Pi^{ijk}\mu_4 \partial_a f A_i B_j C_k
}   
and similarly for $m_4(A,f,B,C)$, $m_4(A,B,f,C)$ and $m_4(A,B,C,f)$.
One realizes that there appear consistency conditions for the
existence of a solution, that are however satisfied once the
relations that we encountered before are satisfied. After all,
the four parameters in $m_4(f,A,B,C)$ remain as free parameters
with the other three four-products given as
\eqb{
   m_4(A,f,B,C)=&\hat\Pi^{ijk} \Big( -\mu_4 A_a B_i C_j +({\textstyle {2\over
       3}}+\mu_2-x_2) A_j B_a C_i\\
     &\phantom{aaaai} +(\gamma+\mu_3) A_i B_j C_a   \Big) \partial_k f \\
    +&\hat\Pi^{ijk}  (-{\textstyle {2\over
       3}}+\gamma-\mu_1+x_2) \partial_a f A_i B_j C_k
}   
and
\eqb{
   m_4(A,B,f,C)=&\hat\Pi^{ijk} \Big(    (-{\textstyle {2\over
       3}}+\gamma-\mu_4-x_1)  A_a B_i C_j +( \mu_1-x_1-x_2) A_j B_a C_i\\
     &\phantom{aaaai} +(\alpha+\gamma+\mu_3+y_2) A_i B_j C_a   \Big) \partial_k f \\
    +&\hat\Pi^{ijk}  (-{\textstyle {2\over
       3}}-\alpha-\mu_2+x_2-y_2) \partial_a f A_i B_j C_k\\[0.2cm]
     m_4(A,B,C,f)=&\hat\Pi^{ijk} \Big(    (-{\textstyle {2\over
       3}}+\alpha+\gamma-\mu_4-x_1)  A_a B_i C_j +( \mu_1-x_1) A_j B_a C_i\\
     &\phantom{aaaai} +({\textstyle {2\over
       3}}+\mu_2+y_2) A_i B_j C_a   \Big) \partial_k f \\
    +&\hat\Pi^{ijk}  (-\alpha-\gamma-\mu_3-y_2) \partial_a f A_i B_j C_k\,. \nonumber
}   
We note that via graded symmetrization the corresponding $\ell_4$-product
is vanishing, being consistent with our findings in section \ref{sec_main}.

Having now a non-trivial $m_4$-product one also has to worry about the
relation ${\cal A}_5$ that for $m_5=0$ contains the order $O(\hat\Theta^2)$ term
\eq{
   {\cal A}_5 &=m_4(m_2\otimes 1^3 - 1\otimes m_2\otimes 1^2 + 1^2\otimes
   m_2\otimes 1-  1^3 \otimes m_2) \\
  &\phantom{=}+ m_2(m_4\otimes 1- 1\otimes m_4) + O(\hat\Theta^3)\,.
}
We have checked that all ten relations of the type  ${\cal
  A}_5(f,g,A,B,C)$ are satisfied. All higher relations are trivially
satisfied up to order $\hat\Theta^2$.

Let us summarize our findings: We have explicitly constructed the
A$_\infty$ algebra up to order $O(\hat\Theta^2)$ that underlies the
non-commutative gauge theory governed by a non-constant and in general
even non-associative star-product. By constructing the higher products
in a step-by-step procedure, we encountered many derivative
$\partial\Theta$-corrections that make the whole algebra and relations
highly non-trivial. At each step, we observed that the A$_\infty$
relation 
under question led to some consistency conditions that were
automatically satisfied once the lower  A$_\infty$
relations were already satisfied. This is very encouraging and makes
us believe that the whole procedure continues also to higher orders
in $\Theta$. Up to the level that we were considering, we found that
in the associative case, a non-constant $\Theta$ induces non-vanishing
higher products of type 
\eq{
               m_2(f,g)\,,\quad m_2(A,g)\, \quad m_3(A,B,g)\,.
}
In the non-associative case a further three and a four-product had to
be introduced
\eq{
     m_2(f,g)\,,\quad m_2(A,g)\,, \quad m_3(A,f,g)\,, \quad
     m_3(A,B,g)\,, \quad m_4(f,A,B,C)\,.
}

In the conclusion of this section we stress that both, for the
consistency of the proposed construction of A$_\infty$ and for the
correct relation to the L$_\infty$, the product $m_2(f,g)$ was taken
to be the Kontsevich star product $f\bullet g$. Up to  second order
in $\Theta$, $\ell_2(f,g)=i[f,g]_\bullet$, coincides with the
``classical'' (quasi)-Poisson bracket, $-\{f,g\}$, which is not the
case of the product $m_2$. It contains ``quantum'' information in the
sense of deformation quantization corrections. A separate question is
whether there exists a ``classical'' L$_\infty$ algebra, where 
the two-product is always simply $\ell_2(f,g)=\{f,g\}$  or whether for
consistency one should necessarily take $\ell_2(f,g)$ as a star
commutator, i.e. construct ``quantum'' L$_\infty$. 


\clearpage
\bibliographystyle{utphys}
\bibliography{references}  


\end{document}